\documentclass[aps,pra,twocolumn,superscriptaddress]{revtex4-2}

\usepackage{url}
\usepackage[utf8]{inputenc}
\usepackage[T1]{fontenc}

\usepackage[normalem]{ulem}
\usepackage{color}
\usepackage{mathtools}
\usepackage{hyperref}
\usepackage{xfrac}
\usepackage{CJKutf8}

\usepackage{empheq}
\usepackage{cleveref}
\usepackage{graphicx}
\usepackage{float}
\restylefloat{table}

\usepackage{ragged2e}

\usepackage{amsmath,amssymb,amsxtra,amstext,amsfonts,dsfont}
\usepackage{multirow}
\usepackage{array}
\usepackage{upgreek}
\usepackage{mathrsfs}
\usepackage{bbold}
\usepackage{bm}

\usepackage{siunitx}  %newer version than SIunit,use \SI{}{} now instead of \unit{}{}
\usepackage{soul,xcolor}
\setstcolor{red}

\newcommand{\bra}[1]{\ensuremath{\langle{#1}|}}
\newcommand{\ket}[1]{\ensuremath{|{#1}\rangle}}
\newcommand{\bracket}[2]{\ensuremath{\langle{#1}|{#2}\rangle}}
\newcommand{\abs}[1]{\ensuremath{\left|{#1}\right|}}

\newcommand{\Hqd}{\hat H_{\text{qd}}}

\newcommand{\vect}[1]{{\bm{#1}}}			% vectors are bold and italic
\newcolumntype{C}[1]{>{\centering\arraybackslash}p{#1}}

\begin{document}

\title{Deterministic and programmable generation of \\ mechanical multipartite entangled states}

\author{Isabelle Franzen}
\affiliation{Department of Physics, ETH Z\"{u}rich, Z\"{u}rich 8093, Switzerland}

\author{Max-Emanuel Kern}
\affiliation{Department of Physics, ETH Z\"{u}rich, Z\"{u}rich 8093, Switzerland}

\author{Q. Rumman Rahman}
\affiliation{Department of Electrical and Computer Engineering, Princeton University, Princeton, New Jersey 08540, USA}

\author{Igor Kladari\'{c}}
\affiliation{Department of Physics, ETH Z\"{u}rich, Z\"{u}rich 8093, Switzerland}

\author{Stefano Marti}
\affiliation{Department of Physics, ETH Z\"{u}rich, Z\"{u}rich 8093, Switzerland}

\author{Yiwen Chu}
\affiliation{Department of Physics, ETH Z\"{u}rich, Z\"{u}rich 8093, Switzerland}

\author{Zhi-Yuan Wei (\begin{CJK*}{UTF8}{gbsn}魏志远\end{CJK*})}
\email{zywei@umd.edu}
\affiliation{Joint Quantum Institute and Joint Center for Quantum Information and Computer Science, NIST/University of Maryland, College Park, Maryland 20742, USA}

\author{Matteo Fadel}
\email{matteofadel@pku.edu.cn}
\affiliation{Institute of Quantum Science and Technology $\&$ Frontiers Science Center for Nano-Optoelectronics, School of Physics, Peking University, Beijing 100871, China}
\affiliation{Department of Physics, ETH Z\"{u}rich, Z\"{u}rich 8093, Switzerland}

\date{\today}
 
\begin{abstract}
Multimode mechanical resonators offer a large set of densely packed, long-lived bosonic modes within a single compact device, making them attractive for quantum simulation, quantum information processing, and quantum-enhanced sensing.
Many such applications rely on multipartite entanglement, yet deterministic mechanical entanglement has so far been largely limited to pairs of modes.
Here we demonstrate the deterministic and programmable preparation of multipartite entangled states among the phonon modes of a high-overtone bulk acoustic wave resonator coupled to a superconducting transmon qubit.
By sequentially applying optimal-control-engineered isometries between the qubit and successive modes, we generate matrix product states of bond dimension two, including $W$, Greenberger--Horne--Zeilinger (GHZ), and cluster states.
We fully characterize these states via quantum state tomography and certify genuine multipartite entanglement among up to four modes.
We further demonstrate correlated-noise sensing by exploiting the complementary decoherence responses of the stored $W$ and GHZ states, isolating inter-mode phase correlations from local decoherence and placing quantitative bounds on collective dephasing.
Our results establish a scalable approach to preparing multipartite entangled states in mechanical systems, with promising applications in multiparameter sensing and tests of quantum mechanics with massive objects.
\end{abstract}

\maketitle

\begin{figure*}[t]
\centering
\includegraphics[width=175mm]{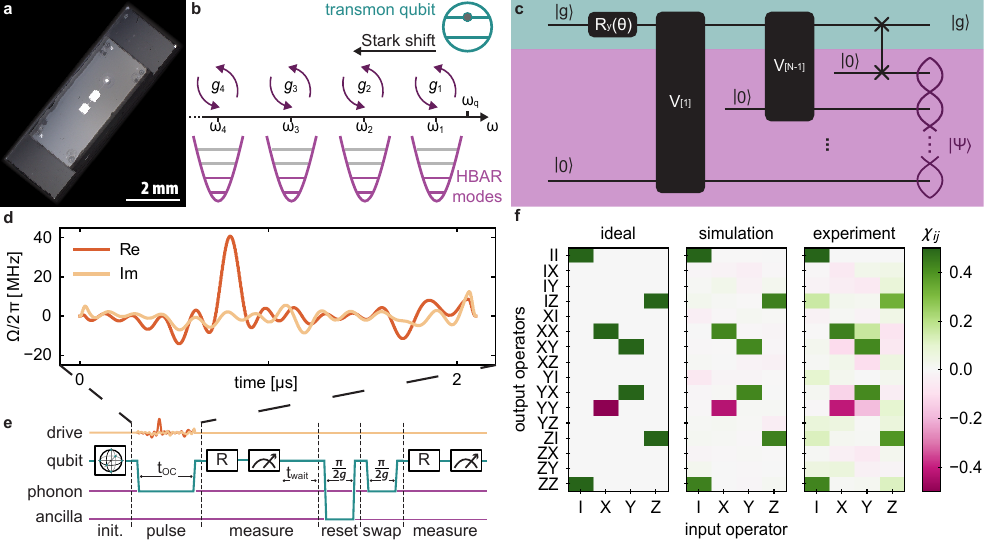}
\caption{\textbf{HBAR device and process tomography of the MPS preparation pulses.} 
\textbf{a)} Micrograph of the transmon-HBAR device and \textbf{b)} its schematic illustration. The transmon can be frequency-shifted to become resonant with the individual phonon modes of the HBAR, to which it couples at rate $g_k$. In this work, we only operate within the $\{\ket{0},\ket{1}\}$ subspace of each phonon mode. 
\textbf{c)} Algorithm to prepare MPS in an HBAR using isometries $\hat V$ and the qubit as an ancilla. 
\textbf{d)} Pulse shape used for implementing the GHZ isometry Eq.~\eqref{eq:isoGHZ}. 
\textbf{e)} Gate sequence used to retrieve process maps. 
\textbf{f)} Process maps of the ideal, simulated, and experimental isometry Eq.~\eqref{eq:isoGHZ} used for GHZ state preparation. See main text for details.
}
\label{fig1}
\end{figure*}

\makeatletter
\let\oldaddcontentsline\addcontentsline
\renewcommand{\addcontentsline}[3]{}
\makeatother

Mechanical degrees of freedom are important building blocks for quantum technologies \cite{ClerkNatReview,Barzanjeh2022}. 
Their bosonic nature provides access to an infinite-dimensional Hilbert space capable of encoding complex quantum states \cite{Ofek2016,Fluhmann2019}, and they can exhibit long coherence times owing to their weak coupling to electromagnetic environments \cite{MacCabe2020,Bozkurt2025,Diamandi2025}.
Interfacing mechanical resonators with superconducting qubits has enabled the preparation and control of nonclassical phonon states in a variety of platforms, including nanomechanical resonators \cite{OConnell2010}, surface acoustic wave devices \cite{GustafssonSci14,Satzinger18,Bienfait2019}, phononic crystal cavities \cite{Arrangoiz2019,wollack_quantum_2022}, and bulk acoustic wave resonators \cite{Chu2018,catSCI23,Marti2024,Yang2024}.
In this context, high-overtone bulk acoustic wave resonators (HBARs) constitute a compelling platform that can offer hundreds of independently addressable long-lived bosonic modes in a compact physical footprint \cite{ChuScience2017,kervinen2018interfacing,gokhale2020epitaxial,vonLupke22,garcia-belles_loss_2026}. 
This unique combination opens new opportunities for quantum simulation \cite{Chen2023,BSpaper}, quantum memories \cite{Hann2019,chamberland2022building}, and quantum sensing \cite{macroPRL,rahman_genuine_2024,LinehanPRD25,Carney21,GoryachevPRD14,LoPRX16,omahen_ultracold_2026}. 
However, to fully exploit the potential of multimode mechanical devices for many of these applications, it is necessary to develop a scalable and programmable approach for the preparation of many-body quantum states across modes.

Many physically relevant many-body quantum states have limited entanglement, in the sense of exhibiting entanglement entropy scaling with the area of a subsystem rather than its volume \cite{EisertRMP10}. 
As a result, one can focus on the subset of states that admit efficient representations in terms of tensor networks \cite{OrusAP14,CiracRMP21}. In one-dimensional systems, this structure is captured by matrix product states (MPS), where the wavefunction is decomposed into a chain of tensors of small bond dimension, providing a compact and efficient description of many-body states \cite{PerezQIC07}.
For this reason, considerable effort has been devoted to developing experimental protocols for the preparation of MPS \cite{schon_sequential_2005,Verstraete2009,Ran2020a,zyadi,malz2023preparation,smith2024constant,wei2025state,tang2026weighted}. 
A particularly powerful approach relies on using a two-level system to sequentially emit a stream of entangled temporal modes of a propagating field \cite{schon_sequential_2005}, as successfully demonstrated in both the optical \cite{SchwartzSCI16,Thomas2022} and microwave \cite{besse_realizing_2020,eichler_exploring_2015,FerreiraNatPhys24,osullivan_deterministic_2025} domains.

Within the class of sequentially generated MPS, three families of multipartite entangled states are of particular importance: $W$, GHZ, and cluster states \cite{horodecki_quantum_2009}. 
$W$ states retain genuine multipartite entanglement even after the loss of individual particles \cite{dur_three_2000}, making them attractive for quantum communication and networking tasks in the presence of loss \cite{Lipinska2018}. 
GHZ states, in contrast, provide the standard resource for fundamental tests of quantum nonlocality \cite{kafatos_going_1989,Mermin1990} and for metrology at the Heisenberg limit \cite{bollinger_optimal_1996}. 
Their fragility, whereby the loss of a single constituent destroys the entanglement, makes them a stringent benchmark for source fidelity.
Cluster states, a paradigmatic class of graph states \cite{BriegelPRL01,hein_multiparty_2004}, are the universal resource underlying measurement-based quantum computation \cite{RaussendorfPRL01,RaussendorfPRA03}, where an entangled substrate is prepared once and computation proceeds through a sequence of adaptive single-qubit measurements.

Previous experiments with mechanical degrees of freedom have shown bipartite entanglement between the motional states of trapped-ion oscillators \cite{jost_entangled_2009}, bulk vibrations in diamonds \cite{LeeSci11}, drumhead oscillators \cite{ockeloen-korppi_stabilized_2018,kotler_direct_2021,mercier_de_lepinay_quantum_2021}, silicon nanobeams \cite{riedinger_remote_2018}, and circuit quantum acoustodynamics (cQAD) devices \cite{wollack_quantum_2022,Qiao23,BSpaper,chou_deterministic_2025,kern_einstein-podolsky-rosen_2026}.
Multipartite mechanical entanglement, however, has so far been detected only in a few experiments, based on multimode squeezing of surface acoustic wave phonons \cite{andersson_squeezing_2022} and dissipative engineering of trapped-ion motional modes \cite{li_programmable_2025}.
Moreover, efficient techniques to prepare such states in a deterministic and programmable manner still have to be developed.

In this work, we demonstrate the deterministic and programmable generation of multipartite entangled states among the gigahertz-frequency phonon modes of a mechanical resonator, in a cQAD platform where a superconducting transmon qubit is coupled to an HBAR.
This is achieved through sequential interactions between the qubit and successive phonon modes, which realize bosonic transformations that prepare arbitrary MPS of bond dimension two.
We characterize the elementary building blocks of this protocol via process tomography, and use them to prepare $W$, GHZ, and cluster states, which we reconstruct via state tomography.
Using entanglement witnesses, we certify genuine multipartite entanglement (GME) among up to four mechanical modes.
Beyond state preparation and entanglement certification, we demonstrate a correlated-noise sensing protocol using the stored mechanical states. Exploiting the complementary sensitivity of $W$ and GHZ states to relative and collective phase fluctuations, we isolate inter-mode correlations in the phonon dephasing from local decoherence, without requiring calibration of the single-mode decay rates. 
This allows us to bound the collective-dephasing rate to over an order of magnitude below the phonon energy-decay rates, an important benchmark for multimode quantum memories and error correction~\cite{Klesse2005,Wilen2021}.
%This allows us to place a quantitative upper bound on collective dephasing, providing an important benchmark for multimode quantum memories and error correction~\cite{Klesse2005,Wilen2021}.
% Since the prepared states remain stored in the resonator, we can follow their free evolution and use them to probe the noise affecting the device.
% Exploiting the complementary sensitivity of $W$ and GHZ states to relative and collective phase fluctuations, we isolate inter-mode correlations in the phonon dephasing from local decoherence, without requiring calibration of the single-mode decay rates.
% We find no evidence of correlated noise, and bound the collective-dephasing rate to over an order of magnitude below the phonon energy-decay rates, an important benchmark for multimode quantum memories and error correction~\cite{Klesse2005,Wilen2021}.
Our results establish a scalable route toward deterministically preparing broad classes of multipartite entangled states in mechanical systems, with direct applications to quantum simulation \cite{foss-feig_holographic_2021,zhang_sequential_2024}, multiparameter quantum sensing \cite{ProctorPRL18,ShettellPRL20,FriisNJP17}, and tests of quantum mechanics in the massive regime \cite{macroPRL,catSCI23}.

\section{The system}
The cQAD device used in this work consists of a transmon qubit flip-chip bonded to an HBAR \cite{vonLupke22}, see Fig.~\ref{fig1}a.

The qubit has a frequency $\omega_q/2\pi \approx \SI{5}{GHz}$, which can be tuned via a far-off-resonant AC Stark shift drive.
At its idle frequency $\omega_{q, \mathrm{idle}}/2\pi = \SI{5.021}{GHz}$, the qubit has an energy relaxation time $T_{1,q}=\SI{28(1)}{\micro s}$, Ramsey decoherence time $T_{2,q}^\ast=\SI{15(1)}{\micro s}$, and anharmonicity $\alpha/2\pi=\SI{185}{MHz}$.

The HBAR has a longitudinal free spectral range of $\SI{12.7}{MHz}$ and is coupled to the qubit through a piezoelectric transducer made of aluminum nitride that mediates a Jaynes--Cummings (JC) interaction. 
The system Hamiltonian reads
\begin{align}
    \hat H/\hbar &= \omega_q \hat q^\dagger \hat q - \dfrac{\alpha}{2} {\hat q^{\dagger 2}} \hat q^2 \notag\\
    &\; + \sum_k \left[ \omega_k \hat a_k^\dagger \hat a_k + g_k (\hat q \hat a_k^\dagger + \hat q^\dagger \hat a_k) \right] + \Hqd/\hbar \;, \label{eq:fullH}
\end{align}
where $\hat q$ and $\hat a_k$ are the bosonic annihilation operators for the qubit and the phonon modes $k \in \{1, 2, 3, 4\}$, respectively, and $g_k$ is the coupling strength. 
To good approximation we find $g_k/2\pi = \SI{273}{kHz}$, independently of $k$.
The term $\Hqd/\hbar=\Omega(t) e^{-i\omega_d t} \hat q^\dagger + \text{h.c.}$ describes a microwave drive at frequency $\omega_{d}$ and time-dependent complex amplitude $\Omega(t)$ applied to the qubit.
This drive will be used to implement qubit-phonon operations as described in the following.
The four phonon modes we consider in this work have frequencies $\omega_k/2\pi = \{\SI{5.0232}{GHz}, \allowbreak \SI{5.0359}{GHz}, \SI{5.0486}{GHz}, \SI{5.0613}{GHz}\}$, energy relaxation times $T_{1,k}=\{\SI{75(5)}{\micro s}, \SI{82(2)}{\micro s}, \SI{57(7)}{\micro s}, \SI{117(7)}{\micro s}\}$ and Ramsey decoherence times $T_{2,k}^{\ast} = \{\SI{149(8)}{\micro s}, \SI{159(6)}{\micro s}, \allowbreak \SI{112(18)}{\micro s}, \SI{216(14)}{\micro s}\}$.

\section{Sequential state-preparation protocol}
Our goal is to demonstrate a scalable and programmable approach for preparing many-body entangled states $\ket{\Psi}$ of phononic modes. 
We focus here on states encoded in the two lowest Fock states of each mode, $\{\ket{0}_k, \ket{1}_k \}$, which we refer to as the computational subspace, and discuss the generalization to larger subspaces later. 
The state preparation approach we consider consists of sequential interactions with the qubit~\cite{schon_sequential_2005}, as illustrated in Fig.~\ref{fig1}c. 
Specifically, we sequentially implement a set of isometries $\{\hat V_{[i]} \}$ acting as maps between the qubit $\{\ket{g}_q,\ket{e}_q\}$ states and (the computational subspace of) each phonon mode. For example, to prepare GHZ states, we are interested in the isometry defined by
\begin{equation}\label{eq:isoGHZ}
\hat{V}:\; \ket{g}_q \ket{0}_k \rightarrow \ket{g}_q \ket{0}_k, \quad \ket{e}_q \ket{0}_k \rightarrow \ket{e}_q \ket{1}_k.
\end{equation}
Note that, although similar to the action of a CNOT gate, here we do not specify the mapping of the other basis states $\ket{g}_q \ket{1}_k$ and $\ket{e}_q \ket{1}_k$. 
This crucial reduction of constraints allows for implementing isometries considerably faster than unitaries~\cite{SM}. 
As shown in Fig.~\ref{fig1}c, after having implemented $N-1$ isometries we perform an iSWAP operation between the qubit state and a phonon mode, which results in the preparation of an $N$-partite phonon state with the qubit disentangled from it.

To experimentally realize a desired isometry ${\hat V}$, we introduce a protocol in which the qubit is tuned into resonance with the target phonon mode, while simultaneously applying a microwave drive also at the same frequency. 
This drive has a complex, time-dependent amplitude $\Omega(t)$, obtained from an optimal control (OC) algorithm \cite{zycq,zyryd,SM}, see e.g. Fig.~\ref{fig1}d.
By suitably shaping $\Omega(t)$, we can realize an arbitrary isometry between the qubit and the phonon's computational subspace, thereby enabling the preparation of arbitrary matrix product states of phononic modes with bond dimension $D=2$. 
In the following, we will investigate specific instances of such states, namely $W$, GHZ, and cluster states.

In the ideal case of a closed system, our simulations predict that the isometry Eq.~\eqref{eq:isoGHZ} can be realized with $>99\,\%$ fidelity in a time $t \sim \pi/g$. Its experimental implementation, however, is unavoidably affected by decoherence.
To benchmark the fidelity of the implemented operation, we perform process tomography~\cite{SM}.
We prepare the qubit in one of the six cardinal points of the Bloch sphere, $\{\ket{g/e},\ket{\pm},\ket{\pm i} \}$, then apply the OC pulse realizing the desired isometry, and finally perform full state tomography of the joint qubit-phonon system, see Fig.~\ref{fig1}e for a sketch of this sequence. 
By repeating this experiment $\sim 10^4$ times for each of the six initial qubit states we are able to reconstruct the process map $\chi$ associated with the OC pulse \cite{SchwartzSCI16,SM}. 
The resulting process map is illustrated in Fig.~\ref{fig1}f, and it shows a fidelity of $90.8 (4)\,\%$ ($89.2 (4)\,\%$ including SPAM errors) compared to the ideal map. This is comparable to the $93.5\,\%$ fidelity of the simulated map, which takes into account decoherence of the system during the OC pulse.

We perform a similar characterization for the qubit-phonon iSWAP operation, which is implemented by letting the qubit and phonon mode interact resonantly according to the JC Hamiltonian for a time $t_\text{swap}=\pi/2g$. Process tomography gives us a fidelity of $96.1 (4)\,\%$ ($94.4 (4)\,\%$ including SPAM errors) for the experimental operation and of $96.5\,\%$ for the simulated one \cite{SM}.

To correct for the SPAM error contribution to a measured gate fidelity, we divide the measured gate fidelity by the identity gate fidelity of $98.2(2)\,\%$ \cite{korotkov2013}. We have verified that the results obtained using this method are consistent with the benchmarking protocol of repeating a gate $N_G$ times and fitting the measured gate fidelities for each repetition number to the exponential function $\mathcal{F}(N_G) = A (\mathcal{F}_{\rm gate})^{N_G} + B$, where $\mathcal{F}_{\rm gate}$ is the gate fidelity excluding SPAM errors and $A$, $B$ are free parameters that account for the latter.

Having demonstrated the realization and characterization of one step of the state preparation protocol shown in Fig.~\ref{fig1}c, in the following we will show the preparation of paradigmatic classes of multipartite entangled states.

\begin{figure*}[t]
\centering
\includegraphics[width=175mm]{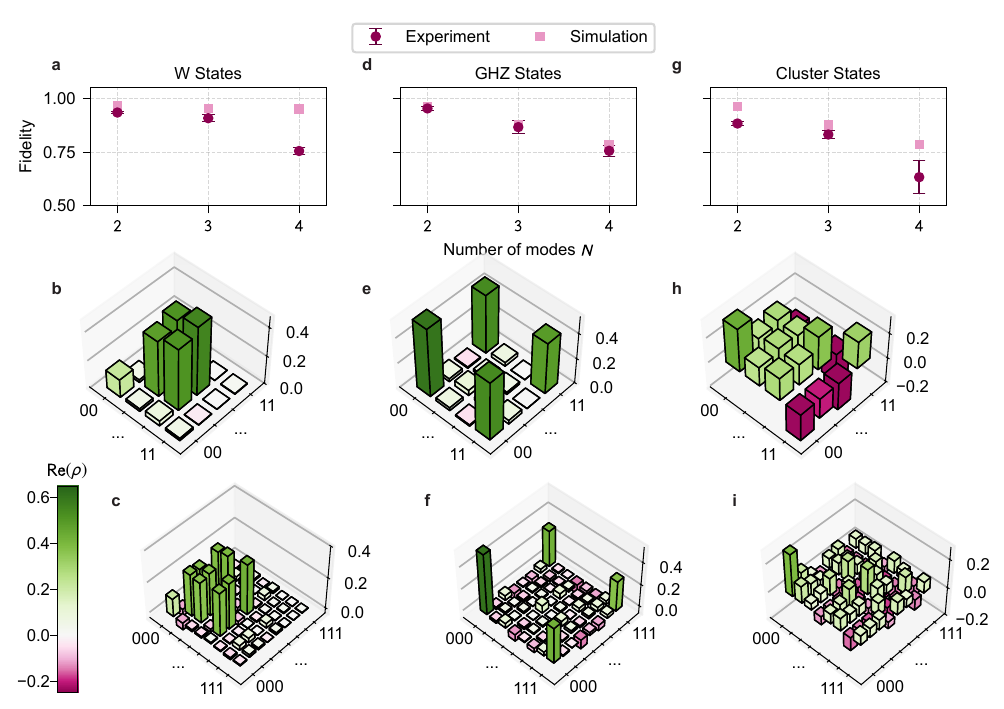}
\caption{\textbf{Fidelities of the prepared $W$, GHZ and cluster states, and density matrices.}
\textbf{a)} Experimental and simulated fidelities for $W$ states prepared on different numbers of modes. 
\textbf{b,c)} Real part of representative density matrices for $W_2$ and $W_3$ states reconstructed from full state tomography.
\textbf{d,e,f)} and \textbf{g,h,i)} show the same for GHZ and cluster states, respectively.
Simulations include decoherence during the state-preparation sequence but not the sequential tomographic readout.
For $W$ states the decrease of the measured fidelities thus originates mainly from the readout, while for GHZ and cluster states the decrease reflects decoherence during the optimal-control pulses (see main text).
}
\label{fig2}
\end{figure*}

\section{State preparation and tomography}

The protocol we use to prepare phononic many-body states is illustrated in Fig.~\ref{fig1}c. First, the qubit is initialized in a desired state $\ket{\phi_0}_q$ by a single-qubit operation $\hat R_y(\theta)$.
Then, an isometry $\hat V_{[1]}$ between the qubit and the first phonon mode is implemented by tuning the two systems into resonance and sending the corresponding OC pulse, as discussed in detail in the previous section.
This step is repeated on each of the phonon modes participating in the many-body state we want to prepare, with in general different isometries $\hat V_{[i]}$.
Finally, the state preparation sequence is concluded by swapping the qubit state to the last phonon mode, which disentangles the qubit from the mechanics.
In the following, we describe how this general scheme is specialized to prepare three paradigmatic families of multipartite entangled states, namely $W$, GHZ, and linear cluster states, which differ in the initial qubit state and in the choice of isometries.

\textbf{$W$ states.}
The simplest case to consider is the preparation of $W$ states, namely states $\ket{W_N}=(\ket{10\dots0}+\ket{01\dots0}+\dots+\ket{0\dots01})/\sqrt{N}$ where one excitation is delocalized over $N$ modes.
The protocol for their preparation relies on initializing the qubit in the excited state $\ket{\phi_0}_q=\ket{e}$, followed by the implementation of isometries $\hat{V}_{[i]}$, for $i=1,\dots,N-1$, effectively equivalent to an $\text{iSWAP}(\theta_i)$ gate with angles $\theta_i = 2 \arcsin\left[(N-i+1)^{-1/2}\right]$.
These partial-iSWAP isometries do not require an OC pulse: with the qubit drive switched off, $\Omega(t)=0$, the qubit and the phonon mode simply evolve on resonance under the JC Hamiltonian for a time $t_i = \theta_i/2g$.

\textbf{GHZ states.}
The preparation of GHZ states $\ket{\text{GHZ}_N}=(\ket{0\dots0}+\ket{1\dots1})/\sqrt{2}$ begins with initializing the qubit in $\ket{\phi_0}_q=(\ket{g}+\ket{e})/\sqrt{2}$, followed by the implementation of the isometry $\hat{V}$ defined in Eq.~\eqref{eq:isoGHZ} on each of the first $N-1$ modes. Finally, the iSWAP operation between the qubit and the $N$th mode results in an $N$-partite GHZ state between the phonon modes.
We note that the duration, and thus infidelity, of this state preparation sequence can be further reduced by replacing the first isometry with a gate sequence that prepares the state $(\ket{g0}+\ket{e1})/\sqrt{2}$ between qubit and first phonon mode. 
This is achieved by first preparing the two systems in a $W$ state, as discussed in the previous paragraph, and then applying a local $\pi$-pulse to the qubit.

\textbf{Cluster states.}
The preparation of linear cluster states begins by initializing the qubit in $\ket{\phi_0}_q=(\ket{g}+\ket{e})/\sqrt{2}$, followed by the isometry that maps $\ket{g}_q \ket{0}_k \rightarrow \frac{1}{\sqrt{2}}(\ket{g}_q + \ket{e}_q) \ket{0}_k$ and $\ket{e}_q \ket{0}_k \rightarrow \frac{1}{\sqrt{2}}(\ket{g}_q - \ket{e}_q) \ket{1}_k$. This is equivalent to using the isometry Eq.~\eqref{eq:isoGHZ} followed by a Hadamard gate applied to the qubit. 
As for the GHZ case, the duration of this protocol can also be reduced by replacing the first isometry with a set of gates that prepare a qubit-phonon cluster state. 

\vspace{2mm}
In order to characterize the prepared many-body states, we use the qubit to perform full state tomography \cite{BSpaper}.
For this, we swap the state of phonon mode $k$ to the qubit initialized in $\ket{g}$, apply a single-qubit unitary to set the measurement axis, and then perform a single-shot readout of the qubit state with a $\SI{4}{\micro s}$-long pulse on the readout resonator.
We then wait $\SI{2}{\micro s}$ for the latter resonator to return to its ground state, and swap the qubit state back to mode $k$, which resets the qubit to $\ket{g}$ while the already-measured mode takes up the residual excitation.
Each measured mode therefore adds about $\SI{6}{\micro s}$ to the sequence, during which the modes still awaiting readout decay.
Within a single experimental shot, this sequence is applied sequentially to all modes, yielding one joint measurement outcome for all $N$ modes. The full shot is then repeated for all combinations of qubit measurement axes, so as to estimate the $N$-body correlators from which the density matrix is reconstructed.

For all prepared states, we quantify the agreement between the reconstructed density matrix $\rho_{\text{exp}}$ and the ideal target state $\ket{\Psi}$ through the fidelity $\mathcal{F}=\sqrt{\bra{\Psi}\rho_{\text{exp}}\ket{\Psi}}$, after compensating, via local virtual rotations, the relative phases between the branches of the superposition that arise from the different mode frequencies and from the qubit-phonon interactions.
We show in Fig.~\ref{fig2} the resulting fidelities for $W$, GHZ, and cluster states, together with the real parts of representative reconstructed density matrices for $N=2,3$.
We compare the measured fidelities with a simulation of the state-preparation sequence, which includes decoherence of the qubit and phonon modes but not the tomographic readout.
This comparison shows that the dominant sources of infidelity differ between the state families.
For $W$ states, the partial-iSWAP operations are short, and the simulated fidelities remain above $0.95$ for all $N\leq 4$. The decrease of the measured fidelities therefore originates from errors outside the preparation, which we attribute to the sequential tomographic readout, whose duration and number of operations grow with $N$.
For GHZ and cluster states, the preparation relies on considerably longer OC pulses, and decoherence during these pulses becomes the main limitation, as reflected by the decrease of the simulated fidelities with $N$. 
While the measured GHZ fidelities agree with the simulation within the experimental uncertainty, the cluster-state fidelities lie systematically below it, which we attribute to their more complex structure and correspondingly higher sensitivity to residual preparation errors not included in the simulation.

\begin{figure*}[t]
\centering
\includegraphics[width=175mm]{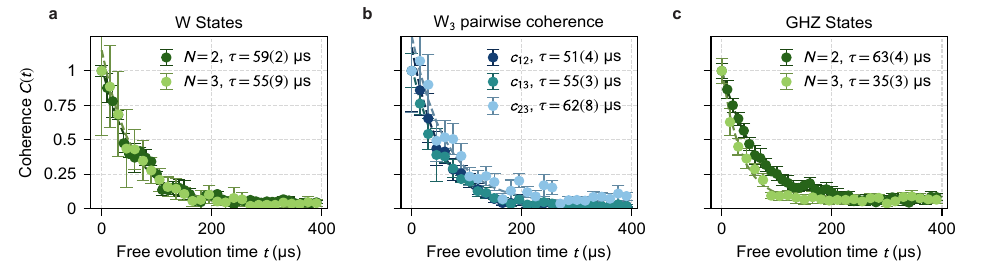}
\caption{
\textbf{Free evolution of $W$ and GHZ coherences.}
\textbf{a)} Coherence $\mathcal{C}(t)$, normalized to its value at $t=0$, of $W$ states with $N=2,3$ modes as a function of the free evolution time $t$. Lines show fits to $A e^{-t/\tau}+B$, with the extracted decay times $\tau$ given in the legend.
\textbf{b)} Pairwise coherences $c_{kl}(t)$ of the $W_3$ state, fitted in the same way.
\textbf{c)} Same as panel \textbf{a}, but for GHZ states. Unlike for $W$ states, the GHZ decay rate depends on $N$. This dependence is here quantitatively explained by local decoherence alone, without any contribution from correlated dephasing (see main text for details).
}

\label{fig3}
\end{figure*}

\begin{table}[t]
\caption{\justifying \textbf{GME witnessing results.}
Expectation values of GME witnesses tailored to the state under investigation, without and with filtering (indicated with a prime).
Uncertainties are standard deviations (STD) estimated from independent state preparations. 
For the $N=4$ cluster state, neither witness certifies GME. Negative values reveal GME.
}
\label{tab:GMEw}
\centering
\renewcommand{\arraystretch}{1.3}
\newcommand{\grouplabel}[1]{%
  \multirow{3}{*}{\rotatebox[origin=c]{90}{#1}\hspace{4pt}%
  $\vcenter{\hbox{\rule{\arrayrulewidth}{3\normalbaselineskip}}}$}}
%\begin{tabular}{c@{\hspace{6pt}}C{1.5cm}C{1.5cm}C{2cm}C{2cm}}
\setlength{\tabcolsep}{10pt}
\begin{tabular}{ccccc}
 & $N$ & criterion & $\text{Tr}[\hat{\mathcal{W}}\rho]$ & $\text{Tr}[\hat{\mathcal{W}}'\rho]$ \\
\cline{2-5}
\noalign{\vskip\doublerulesep}
\cline{2-5}
\grouplabel{$W$}
                   & 2 & $\hat{\mathcal{W}}_1$  & $-0.37(1)$ & $-0.46(1)$ \\
                   & 3 & $\hat{\mathcal{W}}_1$  & $-0.16(3)$ & $-0.19(3)$ \\
                   & 4 & $\hat{\mathcal{W}}_2$ & $0.03(2)$ & $-0.09(2)$ \\
\cline{2-5}
\grouplabel{GHZ}
                   & 2 & $\hat{\mathcal{W}}_3$  & $-0.41(2)$ & $-0.42(2)$ \\
                   & 3 & $\hat{\mathcal{W}}_3$  & $-0.25(5)$ & $-0.28(8)$ \\
                   & 4 & $\hat{\mathcal{W}}_3$ & $-0.07(3)$ & $-0.23(10)$ \\
\cline{2-5}
\grouplabel{Cluster}
                   & 2 & $\hat{\mathcal{W}}_3$  & $-0.28(2)$ & $-0.30(2)$ \\
                   & 3 & $\hat{\mathcal{W}}_3$  & $-0.19(3)$ & $-0.28(4)$ \\
                   & 4 & $\hat{\mathcal{W}}_3$  & $0.09(9)$ & $0.08(9)$ \\
\cline{2-5}
\end{tabular}
\end{table}

\section{Genuine multipartite entanglement}
Entanglement in many-body systems can exhibit rich and complex structures. Its strongest form, genuine multipartite entanglement (GME), arises when a state cannot be written as a convex combination of states that are each separable with respect to some bipartition of the parties (biseparable states), implying quantum correlations shared among all parties.
Operationally, GME can be revealed through a witness operator $\hat{\mathcal{W}}$ satisfying $\text{Tr}[\hat{\mathcal{W}}\sigma]\geq 0$ for all biseparable states $\sigma$.
Therefore, observing $\text{Tr}[\hat{\mathcal{W}}\rho]<0$ implies that $\rho$ is GME.

Several GME witnesses are known, each tailored to some target state of interest.
For the case of $W$ states, one possible choice is the operator \cite{bourennane_experimental_2004,haffner_scalable_2005}
\begin{equation}
    \hat{\mathcal{W}}_1(N) = \dfrac{N-1}{N}\mathbb{1} - \ket{W_N}\bra{W_N} \;,
\end{equation}
where $(N-1)/N$ is the maximal overlap of $\ket{W_N}$ with any biseparable state.
With our fidelity convention, this corresponds to saying that observing a state fidelity larger than $\sqrt{(N-1)/N}$ witnesses GME.
According to this condition, our prepared states show GME up to $N=3$, see Fig.~\ref{fig2}a.
For $N=4$, this fidelity threshold becomes too demanding to certify GME from our data alone, and we instead construct a witness that exploits additional information about the state \cite{haffner_scalable_2005}. We define the operator
\begin{equation}
\hat{Q}(N,\beta) = \ket{W_N}\bra{W_N} - \beta \sum_{k=1}^N \ket{\mathrm{BS}_k}\bra{\mathrm{BS}_k},
\end{equation}
where $\ket{\mathrm{BS}_k}=\ket{0}_k\ket{W_{N-1}}$ denotes the biseparable state with no excitation in mode $k$ and a single excitation delocalized over the remaining $N-1$ modes, which has maximal overlap with $\ket{W_N}$, and $\beta\geq0$ is a free parameter.
For a given $\beta$, we compute $\gamma(N,\beta) = \max_{\sigma\in\mathrm{bisep}} \text{Tr}[\hat{Q}(N,\beta)\,\sigma]$, the maximum expectation value of $\hat{Q}(N,\beta)$ over all biseparable states (see the Supplementary Material \cite{SM} for the full construction).
This allows us to construct the GME witness operator
\begin{equation}\label{eq:W_witness}
    \hat{\mathcal{W}}_2(N,\beta) = \gamma(N,\beta)\mathbb{1}_{(2)} - \hat{Q}(N,\beta) \;,
\end{equation}
with $\mathbb{1}_{(2)}$ the identity operator restricted to at most two excitations.
Since $\hat{\mathcal{W}}_2(N,\beta)$ is a valid GME witness for every $\beta\geq0$, we numerically optimize $\beta$ so as to minimize its expectation value on the reconstructed state $\rho$.

The detection capability of a witness $\hat{\mathcal{W}}$ can be further enhanced by applying a local filtering operation $\hat{F}=\hat{A}_1\otimes\cdots\otimes \hat{A}_N$, where each $\hat{A}_k$ is an arbitrary (in general non-unitary) operator acting on mode $k$, and defining the filtered operator $\hat{\mathcal{W}}'=\hat{F}^\dagger \hat{\mathcal{W}}\hat{F}$ \cite{kiesel_experimental_2007,haffner_scalable_2005}.
Since local operations map biseparable states onto (unnormalized) biseparable states, $\text{Tr}[\hat{\mathcal{W}}'\sigma]=\text{Tr}[\hat{\mathcal{W}}\hat{F}\sigma \hat{F}^\dagger]\geq 0$ holds for every biseparable $\sigma$, so that $\hat{\mathcal{W}}'$ is again a valid GME witness.
Observing $\text{Tr}[\hat{\mathcal{W}}'\rho]<0$ therefore certifies GME of $\rho$, and a suitably chosen filter can detect GME in states for which the unfiltered witness fails, by compensating for the specific imperfections of the measured state (see Supplementary Material \cite{SM}).
Since the magnitude of $\text{Tr}[\hat{\mathcal{W}}'\rho]$ depends on the normalization of $\hat{F}$, we normalize the filter such that $\text{Tr}[\hat{F}\rho\hat{F}^\dagger]=1$ and we assess the significance of each violation by its ratio to the statistical uncertainty.
For $W$ states, we optimize $\beta$ and the filter simultaneously, so as to obtain a single witness maximizing the average violation over all experimental runs.
For our 18 experimental preparations of four-partite $W$ states, we obtain an average violation of $\num{-0.09(2)}$ for $\beta=0.30$, where the uncertainty denotes the standard deviation over the experimental preparations. This violation by more than three standard deviations extends our certification of GME in $W$ states to $N=4$.

For $N$-partite graph states $\ket{G_N}$, which include GHZ and cluster states, we use the GME witness operator \cite{TothPRL05}
\begin{equation}
    \hat{\mathcal{W}}_3(N) = \dfrac{1}{2}\mathbb{1} - \ket{G_N}\bra{G_N} \;,
\end{equation}
where $1/2$ is the maximal overlap of a graph state with any biseparable state. 
With our convention, this corresponds to saying that observing a state fidelity larger than $1/\sqrt{2}$ reveals GME.
Applying this criterion to the data shown in Fig.~\ref{fig2}d,g reveals GME up to $N=3$ for both GHZ and cluster states with a significance exceeding three standard deviations.
As for the $W$ states above, the detection can be improved via a local-filtering-optimized witness.
This allows us to certify GME also for $N=4$ GHZ states, while for $N=4$ cluster states the noise is too large to certify GME with either the filtered or unfiltered witness.
A summary of the GME witness violations is presented in Tab.~\ref{tab:GMEw}.

% cite corr noise references, noise sensing, entanglement advantage, ...
% explain meaning, GME useful for correlated noise sensing, demonstrate ent-enhanced noise sensing, ..

\section{State decoherence and correlated noise sensing}
A key feature of our platform, compared to the sequential emission of MPS into propagating optical or microwave modes, is that the prepared states remain stored in long-lived mechanical modes. 
This enables applications in quantum simulation and metrology where the entire many-body state can be further manipulated or left to evolve in time.

We exploit this capability to use the prepared states as probes of correlated noise among the mechanical modes.
Such correlations are of central importance for multimode devices: they cause the superdecoherence of highly entangled states~\cite{Monz2011} and violate the assumption of independent errors that underlies quantum error correction~\cite{Klesse2005,Wilen2021}, yet single-mode measurements cannot distinguish them from local dephasing.
Entangled states are natural probes of this structure, since their coherences depend directly on the relative or total phases of the modes~\cite{Reina2002}, and they can even outperform product states when the noise is close to maximally correlated~\cite{Gardner2025,Wang2026}.
We prepare $W$ and GHZ states of $N=2,3$ modes, let them evolve freely for a time $t$, and perform full state tomography.
From the reconstructed density matrix we compute the coherence normalized to its initial value,
\begin{equation}\label{eq:coherence}
\mathcal{C}(t)=\frac{\sum_{\mathbf{m}\neq\mathbf{n}} M_{\mathbf{mn}}\abs{\rho_{\mathbf{mn}}(t)}}{\sum_{\mathbf{m}\neq\mathbf{n}} M_{\mathbf{mn}}\abs{\rho_{\mathbf{mn}}(0)}}\;,
\end{equation}
where $M_{\mathbf{mn}}=1$ if the off-diagonal element $\rho_{\mathbf{mn}}$ is non-zero in the ideal state and $M_{\mathbf{mn}}=0$ otherwise.
For $W_3$ states we additionally consider the three individual pairwise coherences $c_{kl}(t)=\abs{\rho_{1_k,1_l}(t)}/\abs{\rho_{1_k,1_l}(0)}$, where $1_k$ denotes a single excitation in mode $k$.

To interpret the coherence dynamics, we model the evolution of the state with the master equation
\begin{align}\label{eq:mastereq}
\dot{\rho}_{\bm{\xi}}(t)
& = -i \left[ \sum_{k=1}^{N}\xi_k(t)\,\hat n_k,\rho_{\bm{\xi}}(t) \right] + \sum_{k=1}^{N}\gamma_1^{(k)} \mathcal{D}[\hat a_k]\rho_{\bm{\xi}}(t) \notag \\ 
& \qquad + \sum_{k=1}^{N} 2 \gamma_{\varphi}^{(k)} \mathcal{D}[\hat n_k]\rho_{\bm{\xi}}(t),
\end{align}
where $\mathcal{D}[\hat L]\rho = \hat L\rho\hat L^{\dagger} -\frac{1}{2}\{\hat L^{\dagger}\hat L,\rho\}$, and $\gamma_1^{(k)}$, $\gamma_\varphi^{(k)}$ are the local loss and pure-dephasing rates of mode $k$. 
The classical stochastic detunings $\xi_k(t)$ describe additional phase noise of arbitrary strength, temporal correlations, and inter-mode correlations, with accumulated phases $\Phi_k(t)=\int_0^t\xi_k(t')\,\mathrm{d}t'$. 
Purely collective dephasing corresponds to the special case $\xi_k(t)=\xi(t)$ for all $k$.
For noise that is white on the timescale of the experiment, averaging over $\bm\xi$ yields the correlated-dephasing Lindbladian describing spatially correlated noise~\cite{Reina2002,Wang2026,Seif2022}.

For a coherence $\rho_{\mathbf{mn}}=\bra{\mathbf m}\rho\ket{\mathbf n}$ of a $W$ or GHZ state, with occupation difference $\mathbf d=\mathbf m-\mathbf n$, the stochastic phases factor out of the local dynamics, and for Gaussian phase noise one finds~\cite{SM}
\begin{equation}\label{eq:coh_main}
\frac{\abs{\rho_{\mathbf{mn}}(t)}}{\abs{\rho_{\mathbf{mn}}(0)}}=\exp\Big[-\sum_k d_k^2\lambda_k t-\frac12\sum_{k,l}d_kd_l\,\mathcal{K}_{kl}(t)\Big],
\end{equation}
where $\lambda_k=\gamma_1^{(k)}/2+\gamma_\varphi^{(k)}$ is the local coherence decay rate of mode $k$, and $\mathcal{K}_{kl}(t)=\langle\delta\Phi_k(t)\,\delta\Phi_l(t)\rangle$, with $\delta\Phi_k=\Phi_k-\langle\Phi_k\rangle$, is the covariance matrix of the accumulated phases.
The diagonal elements $\mathcal{K}_{kk}$ act as additional local dephasing, while only the off-diagonal elements $\mathcal{K}_{k\neq l}$ reflect noise correlations between modes.

The two families of states probe this covariance in complementary ways. 
Every coherence of $W_N$ connects states with the same total excitation number ($\sum_k d_k=0$), and is therefore sensitive only to phase differences $\Phi_k-\Phi_l$. 
In particular, $W$ states are decoherence-free under collective dephasing \cite{Jeske2014}. 
In contrast, the two branches of GHZ$_N$ differ by $N$ excitations and probe the total phase $\sum_k\Phi_k$, making them maximally sensitive to common-mode noise. 
Comparing the two on the same set of modes cancels all local contributions exactly. For $N=2$, we find
\begin{equation}\label{eq:chi2}
\mathcal{R}_2(t)\equiv\frac{\mathcal{C}_{\mathrm{GHZ}_2}(t)}{\mathcal{C}_{W_2}(t)}=\exp\left[-2\,\mathcal{K}_{12}(t)\right],
\end{equation}
while for $N=3$ the square root of the product of the three pairwise $W_3$ coherences reproduces exactly the local decay of the GHZ$_3$ coherence, giving
\begin{equation}\label{eq:chi3}
\mathcal{R}_3(t)\equiv\frac{\mathcal{C}_{\mathrm{GHZ}_3}(t)}{\sqrt{c_{12}(t)\,c_{13}(t)\,c_{23}(t)}}=\exp\Big[-\tfrac32\sum_{k<l}\mathcal{K}_{kl}(t)\Big].
\end{equation}
Equations~\eqref{eq:chi2} and \eqref{eq:chi3} hold for arbitrary, mode-dependent loss and local dephasing rates. Moreover, for statistically independent phase noise on each mode, $\mathcal{R}_2=\mathcal{R}_3=1$ for arbitrary noise statistics and spectra, since the single-mode contributions cancel exactly in both ratios.
They therefore require no independent calibration of the loss and pure-dephasing rates, and deviate from unity only in the presence of inter-mode phase correlations. 
For purely collective noise, $\mathcal{K}_{kl}=\langle\delta\Phi(t)^2\rangle$ and $\mathcal{R}_N(t)=\exp[-N^2\langle\delta\Phi(t)^2\rangle/2]$, recovering the characteristic $N^2$ scaling of GHZ superdecoherence \cite{Monz2011}.

Experimentally, all measured coherences are well described by exponential decays $A\,e^{-t/\tau}+B$, where the offset $B$ accounts for the positive bias of $\abs{\rho_{\mathbf{mn}}}$ when estimated from noisy data. 
Fitting our data with this model gives the decay times $\tau$ shown in Fig.~\ref{fig3}.
The pairwise decay times $\tau_{kl}$ of the $W_3$ coherences $c_{kl}$ are mutually consistent, and their mean decay rate is in agreement with the directly fitted $\tau_{W_3}$.
Within this exponential description, the ratios decay as $\mathcal{R}_N(t)=e^{-\Gamma_N t}$, with $\Gamma_2=\tau_{\mathrm{GHZ}_2}^{-1}-\tau_{W_2}^{-1}$ and $\Gamma_3=\tau_{\mathrm{GHZ}_3}^{-1}-\tfrac12\sum_{k<l}\tau_{kl}^{-1}$.
Without any free parameter, the pairwise lifetimes predict a purely local GHZ$_3$ lifetime of $\tau_{\mathrm{loc}}=\big(\tfrac12\sum_{k<l}\tau_{kl}^{-1}\big)^{-1}=\SI{37(2)}{\micro s}$, which accounts for $94(10)\,\%$ of the measured GHZ$_3$ decay rate.
The faster decay of GHZ$_3$ relative to $W_3$ is thus consistent, within uncertainty, with local decoherence alone, whose contribution grows with the number of excited modes.
We find $\Gamma_2=\SI{-1.0(1.2)}{ms^{-1}}$ and $\Gamma_3=\SI{1.8(3.0)}{ms^{-1}}$, both compatible with zero, so that no inter-mode phase correlations are resolved.

Since $\mathcal{R}_N$ depends only on the inter-mode phase covariances, these results bound correlated phase noise without assumptions on its spatial structure.
If the covariances grow linearly in time on the timescale of the experiment, $\mathcal{K}_{kl}(t)=2D_{kl}t$, as suggested by the observed exponential decays, the off-diagonal elements $D_{kl}$ act as correlated dephasing rates. Equations~\eqref{eq:chi2} and \eqref{eq:chi3} then give $D_{12}=\Gamma_2/4<\SI{0.26}{ms^{-1}}$ and $\sum_{k<l}D_{kl}=\Gamma_3/3<\SI{2.3}{ms^{-1}}$ at the $95\,\%$ confidence level (one-sided).
For purely collective noise, all $D_{kl}$ are equal to $\Gamma_c/2$, with $\Gamma_c$ the collective-dephasing rate defined by $\langle\delta\Phi(t)^2\rangle=\Gamma_c t$. 
The two data sets then measure the same parameter, giving $\Gamma_c=\Gamma_2/2=\SI{-0.5(6)}{ms^{-1}}$ for $N=2$ and $\Gamma_c=2\Gamma_3/9=\SI{0.4(7)}{ms^{-1}}$ for $N=3$, which agree within one standard deviation.
Although the negative $N=2$ value is unphysical, it is consistent with a statistical fluctuation.
The inverse-variance weighted mean, $\bar\Gamma_c=\SI{-0.1(4)}{ms^{-1}}$, is compatible with zero and bounds the collective-dephasing rate to $\Gamma_c<\SI{0.7}{ms^{-1}}$ at the $95\,\%$ confidence level (one-sided).
Collective noise therefore contributes at most $\Gamma_c/2<\SI{0.3}{ms^{-1}}$ to the pure-dephasing rate of each mode, corresponding to a collective dephasing time exceeding $\SI{3}{ms}$, which is up to 50 times longer than the phonon energy-relaxation times. 
A summary of estimated parameters is given in Tab.~\ref{tab:corrnoise}.

We note that the measured $W$-state lifetimes are slightly shorter than predicted from the independently calibrated single-mode decay times, possibly because of drifts between calibration and measurement.
This discrepancy does not affect our conclusions, since local contributions cancel exactly in Eqs.~\eqref{eq:chi2} and \eqref{eq:chi3}. 
These ratios have the advantage of being insensitive to such drifts as long as the data are acquired under the same conditions.
Overall, the coherence dynamics of both $W$ and GHZ states are consistent with independent local decoherence, with no statistically significant contribution from correlated dephasing among the phonon modes.

\begin{table}[t]
\caption{\justifying \textbf{Correlated-noise sensing.}
Quantities derived from the fitted decay times of Fig.~\ref{fig3}.
Uncertainties are standard deviations propagated from the fit uncertainties.
Upper limits are one-sided at $95\,\%$ confidence.
The collective-dephasing rate $\Gamma_c$ assumes purely collective noise, and $\bar\Gamma_c$ is the inverse-variance weighted mean of the $N=2$ and $N=3$ estimates.
}
\label{tab:corrnoise}
\centering
\renewcommand{\arraystretch}{1.3}
%\begin{tabular}{C{3.4cm}C{2.2cm}C{2.2cm}}
\setlength{\tabcolsep}{6pt}
\begin{tabular}{ccc}
Quantity & Value & $95\,\%$ limit \\
\hline
\noalign{\vskip\doublerulesep}
\hline
$\Gamma_2=\tau_{\mathrm{GHZ}_2}^{-1}-\tau_{W_2}^{-1}$                                   & $\SI{-1.0(1.2)}{ms^{-1}}$ & \\
$D_{12}=\Gamma_2/4$                          & $\SI{-0.24(30)}{ms^{-1}}$ & $<\SI{0.26}{ms^{-1}}$ \\
\hline
$\Gamma_3=\tau_{\mathrm{GHZ}_3}^{-1}-\tau_{\mathrm{loc}}^{-1}$                                   & $\SI{1.8(3.0)}{ms^{-1}}$  & \\
$\tau_{\mathrm{loc}}=\big(\frac12\sum_{k<l}\tau_{kl}^{-1}\big)^{-1}$                        & $\SI{37(2)}{\micro s}$    & \\
%$\tau_{\mathrm{GHZ}_3}/\tau_{\mathrm{loc}}$  & $\SI{94(10)}{\percent}$   & \\
$\sum_{k<l}D_{kl}=\Gamma_3/3$                & $\SI{0.6(1.0)}{ms^{-1}}$  & $<\SI{2.3}{ms^{-1}}$ \\
\hline
$\Gamma_c=\Gamma_2/2$                        & $\SI{-0.5(6)}{ms^{-1}}$   & \\
$\Gamma_c=2\Gamma_3/9$                       & $\SI{0.4(7)}{ms^{-1}}$    & \\
$\bar\Gamma_c$                               & $\SI{-0.1(4)}{ms^{-1}}$   & $<\SI{0.7}{ms^{-1}}$ \\
\hline
\end{tabular}
\end{table}

\section{Conclusions and outlook}
We have demonstrated a scalable and programmable protocol for the deterministic preparation of multipartite entangled states in a multimode mechanical resonator.
In our approach, a transmon qubit interacts sequentially with successive phonon modes of an HBAR, and during each interaction an optimal-control pulse realizes a qubit-phonon isometry, the building block for preparing matrix product states of bond dimension two.
We used this protocol to prepare mechanical $W$, GHZ, and cluster states, and certified GME among up to four modes.

The number of modes for which GME can be certified is currently limited by decoherence and errors during both the state preparation and the sequential tomographic readout, whose durations grow with $N$.
A larger qubit-phonon coupling $g$, which shortens the preparation sequence, longer coherence times, faster qubit readout, and characterization schemes requiring only partial information about the state~\cite{frerot_probing_2023} should therefore allow entanglement to be certified among more modes.

Since the prepared states remain stored in the mechanical modes, we further tracked their free evolution in a sensing application, using the complementary sensitivity of $W$ and GHZ states to correlated phase noise to probe the spatial correlations of the phonon dephasing without calibrating the single-mode decay rates.
For both $N=2$ and $N=3$, the inferred collective-dephasing rate is compatible with zero, bounding it to $\Gamma_c<\SI{0.7}{ms^{-1}}$ at the $95\,\%$ confidence level.
The decoherence is thus dominated by local, uncorrelated processes, which, together with the measured Ramsey times $T_2^\ast\simeq 2T_1$, can be attributed to amplitude damping.

Storing the states also means that the qubit is not restricted to a single pass over the modes, but can interact repeatedly with any of them.
Revisiting already entangled modes would establish bonds between arbitrary pairs of modes, analogously to time-delayed feedback in photonic sources \cite{FerreiraNatPhys24}, enabling multipartite entangled states with graph connectivity beyond the one-dimensional, bond-dimension-two case explored here.
These include two-dimensional cluster states, a universal resource for measurement-based quantum computation \cite{RaussendorfPRL01}, and other tensor-network states relevant for multiparameter quantum metrology \cite{ShettellPRL20,FriisNJP17}.
Looking further ahead, extending the sequential-isometry approach beyond the two-level subspace of each mode offers a route toward entangled states of higher-dimensional bosonic modes, including continuous-variable cluster states \cite{GuPRA09}, broadening the applications of HBAR devices.

\vspace{5mm}

\vspace{2mm}
{\centering\textbf{Acknowledgments}\\}
\vspace{2mm}
\noindent The authors thank G\'eza T\'oth for useful discussions on the entanglement witnesses. % and feedback on the manuscript, and ... for contributing to ... 
ZYW was supported in part by the NSF STAQ program (awards No.~1818914 and No.~2325080).
MF was supported by the Swiss National Science Foundation Ambizione Grant No. 208886, Spark Grant No. 237484, and The Branco Weiss Fellowship -- Society in Science, administered by the ETH Z\"{u}rich.

%\vspace{2mm}
%{\centering\textbf{Author contributions}\\}
%\noindent ZYW and MF conceived the experiments. ... and MF wrote experiment control sequences. ... performed measurements and analysed the data. ... and MF performed numerical simulations of the experiments. ... and MF derived theoretical models. YC, ZYW and MF supervised the work. ... and MF wrote the manuscript with feedback from all authors.

%\vspace{2mm}
%{\centering\textbf{Data and code availability}\\}
%\noindent \new{Raw data and analysis scripts are available on Zenodo with ID \href{ https://zenodo.org/records/...}{...}}. Additional material is available from the corresponding author on reasonable request.\\

\clearpage
\newpage

\renewcommand{\addcontentsline}[3]{}% Make 
\bibliography{mybib}
\let\addcontentsline\oldaddcontentsline% Restore \addcontentsline

\clearpage
\newpage

\renewcommand{\thetable}{S\arabic{table}}  
\renewcommand{\thepage}{S\arabic{page}}  
\renewcommand{\thefigure}{S\arabic{figure}}
\renewcommand{\theHfigure}{S\arabic{figure}}
\renewcommand{\theequation}{S\arabic{equation}}
\setcounter{page}{1}
\setcounter{figure}{0}
\setcounter{table}{0}
\setcounter{section}{0}
\setcounter{equation}{0}

%\resetlinenumber
\widetext

{\centering\textbf{\Large Supplementary Material for} \\} 
{\centering\textbf{\Large ``Deterministic and programmable generation of \\ mechanical multipartite entangled states''\\}} 
\normalsize
\vspace{.3cm}
%{\centering ...$^\ast$, ... and Matteo Fadel\\
%\textit{Department of Physics, ETH Z\"{u}rich, 8093 Z\"{u}rich, Switzerland}\\
%\textit{Quantum Center, ETH Z\"{u}rich, 8093 Z\"{u}rich, Switzerland}\\
%}

\vspace{20mm}

\suppressfloats
\tableofcontents

%\vspace{10mm}
\clearpage
\newpage

\section{Sequential MPS preparation in cQAD with optimal control}
\label{sec:MPS}

We implement the sequential state-preparation protocol~\cite{schon_sequential_2005} using quantum optimal-control (OC) pulses adapted from Refs.~\cite{zyryd,zycq} and based on the gradient-optimization approach of GOAT~\cite{Machnes2018}. The protocol uses the transmon qubit as an ancilla that interacts sequentially with phonon modes initially prepared in the vacuum.

\subsection{Sequential generation of phononic MPS}

For each phonon mode $\hat a_k$, we define the local computational subspace
\begin{equation}
    \mathcal H_{k,2}=\operatorname{span}\!\left\{\ket{0}_{k},\ket{1}_{k}\right\}.
\end{equation}
Before its interaction with the qubit, every mode is in
$\ket{0}_{k}$. The physically populated input subspace in round $k$ is
\begin{equation}
    \mathcal H_{\rm in}^{(k)}
    =
    \mathcal H_q\otimes
    \operatorname{span}\!\left\{\ket{0}_{k}\right\}.
\end{equation}
The desired operation is an isometry
\begin{equation}
    \hat V_{[k]}:
    \mathcal H_{\rm in}^{(k)}
    \longrightarrow
    \mathcal H_q\otimes\mathcal H_{k,2},
\end{equation}
whose action on an arbitrary ancilla state $\ket{\varphi}_q$ is
\begin{equation}\label{unit_iso}
    \hat V_{[k]}\!\left(
    \ket{\varphi}_q\ket{0}_{k}
    \right)
    =
    \sum_{j=0}^{1}
    \left(V_{[k]}^j\ket{\varphi}_q\right)
    \ket{j}_{k}.
\end{equation}
Here $V_{[k]}^j$ acts on the qubit Hilbert space. Preservation of inner products by $\hat V_{[k]}$ requires
\begin{equation}
    \sum_{j=0}^{1}V_{[k]}^{j\dagger}V_{[k]}^j=\mathbb{1}_q.
\end{equation}

Starting from the ancilla state $\ket{\varphi_I}_q$, the joint state after applying isometries to the first $k$ modes is
\begin{equation}
    \ket{\Psi_k}_{q,1\ldots k}
    =
    \sum_{j_1,\ldots,j_k=0}^{1}
    V_{[k]}^{j_k}\cdots V_{[1]}^{j_1}
    \ket{\varphi_I}_q
    \otimes
    \ket{j_1\cdots j_k}_{1\ldots k}.
\end{equation}
At this stage the qubit remains entangled with the phonon modes and
carries the uncontracted MPS bond index. 
In the experimental protocol, isometries are applied to the first \(N-1\) modes, followed by a resonant state-transfer operation $\hat S_{q,N}$ between the qubit
and the initially empty $N$th mode. Restricted to the physically
populated input subspace, this operation defines the tensors of the
final MPS site according to
\begin{equation}
    V_{[N]}^{j_N}
    \equiv
    {}_{N}\!\bra{j_N}
    \hat S_{q,N}
    \ket{0}_{N},
    \qquad j_N\in\{0,1\}.
\end{equation}
Equivalently,
\begin{equation}
    \hat S_{q,N}
    \left(\ket{\varphi}_q\ket{0}_{N}\right)
    =
    \sum_{j_N=0}^{1}
    \left(V_{[N]}^{j_N}\ket{\varphi}_q\right)
    \ket{j_N}_{N}.
\end{equation}
For ideal complete state transfer, the range of both
$V_{[N]}^0$ and $V_{[N]}^1$ is the one-dimensional subspace spanned
by $\ket{g}_q$. Consequently,
\begin{equation}
    \hat S_{q,N}
    \left(
    \ket{\Psi_{N-1}}_{q,1\ldots {N-1}}
    \ket{0}_{N}
    \right)
    =
    \ket{g}_q\ket{\psi_{\mathrm{MPS}}},
\end{equation}
where the purely phononic state has the standard open-boundary MPS form
\begin{equation}\label{mps_form}
    \ket{\psi_{\mathrm{MPS}}}
    =
    \sum_{j_1,\ldots,j_N=0}^{1}
    \bra{g}_q
    V_{[N]}^{j_N}
    V_{[N-1]}^{j_{N-1}}
    \cdots
    V_{[1]}^{j_1}
    \ket{\varphi_I}_q
    \ket{j_1\cdots j_N}_{1\ldots N}.
\end{equation}
Thus the qubit provides the bond space of dimension $D=2$, while the final resonant state transfer implements the terminating MPS tensor and disentangles the qubit from the phonon modes.

For the GHZ and linear-cluster states, the same isometry can be used in each intermediate round. In the input basis $\{\ket{g}_q\ket{0}_{k},\ket{e}_q\ket{0}_{k}\}$ and the output basis
\begin{equation}
    \mathcal B_{\mathrm{out}}
    =\left\{
    \ket{g}_q\ket{0}_{k},
    \ket{e}_q\ket{0}_{k},
    \ket{g}_q\ket{1}_{k},
    \ket{e}_q\ket{1}_{k}
    \right\},
\end{equation}
the two isometries are
\begin{equation}\label{iso_form}
\begin{aligned}
\hat V_{\rm GHZ}
&=\begin{pmatrix}V_{[k]}^0\\[1mm]V_{[k]}^1\end{pmatrix}
=\begin{pmatrix}
1&0\\
0&0\\
0&0\\
0&1
\end{pmatrix},
&
\qquad \hat V_{\rm cluster}
&=\begin{pmatrix}V_{[k]}^0\\[1mm]V_{[k]}^1\end{pmatrix}
=\frac{1}{\sqrt{2}}
\begin{pmatrix}
1&0\\
1&0\\
0&1\\
0&-1
\end{pmatrix}.
\end{aligned}
\end{equation}
Their action is equivalently
\begin{align}
    \hat V_{\rm GHZ}\ket{g}_q\ket{0}_{k}
    &=\ket{g}_q\ket{0}_{k},
    &
    \hat V_{\rm GHZ}\ket{e}_q\ket{0}_{k}
    &=\ket{e}_q\ket{1}_{k},\\
    \hat V_{\rm cluster}\ket{g}_q\ket{0}_{k}
    &=\frac{\ket{g}_q+\ket{e}_q}{\sqrt{2}}\ket{0}_{k},
    &
    \hat V_{\rm cluster}\ket{e}_q\ket{0}_{k}
    &=\frac{\ket{g}_q-\ket{e}_q}{\sqrt{2}}\ket{1}_{k}.
\end{align}
Together with $\ket{\varphi_I}_q=(\ket{g}_q+\ket{e}_q)/\sqrt{2}$ and the final iSWAP, these maps generate the corresponding phononic GHZ and linear-cluster states. The $W$-state protocol instead uses round-dependent partial-swap isometries, with interaction angles $\theta_i=2\arcsin[(N-i+1)^{-1/2}]$, as described in the main text.

\subsection{Isometry implementation with optimal control}
\label{goat_imple}

The pulse-synthesis calculation models the resonant qubit-phonon system using a two-level qubit. In the interaction frame of the resonant qubit and target phonon mode, the control Hamiltonian is
\begin{equation}\label{H_ctrl}
    H_{\rm int}(t) / \hbar
    = g_k\left(\sigma^-a_k^\dagger+\sigma^+a_k\right)
    + \operatorname{Re}[\Omega(t)]\,\sigma_x
    - \operatorname{Im}[\Omega(t)]\,\sigma_y.
\end{equation}
Here $a_k$ annihilates an excitation in the target phonon mode, $g_k$ is the fixed resonant Jaynes--Cummings exchange coupling, and $\Omega(t)=\Omega_x(t)+i\Omega_y(t)$ is the complex qubit drive. Its real and imaginary parts provide two independently parameterized control quadratures. 
The pulse optimization itself does not include the transmon state $\ket{f}$ or finite transmon anharmonicity. These effects are restored separately when the optimized pulse is evaluated using the multilevel noisy model employed for comparison with experiment.

The relevant input subspace for each isometry is not the full qubit-phonon computational space, because the phonon mode is initially in vacuum. We define
\begin{equation}
    \mathcal H_{\rm in}
    =\operatorname{span}\!\left\{
    \ket{g}_q\ket{0}_k,\ket{e}_q\ket{0}_k
    \right\},
    \qquad D\equiv\dim\mathcal H_{\rm in}=2,
\end{equation}
and the desired output subspace
\begin{equation}
    \mathcal H_{\rm out}
    =\mathcal H_q\otimes
    \operatorname{span}\!\left\{\ket{0}_k,\ket{1}_k\right\},
    \qquad \dim\mathcal H_{\rm out}=4.
\end{equation}
For the OC pulse synthesis, the phonon Hilbert space is truncated to the levels $n=0,\ldots,7$, such that $N_k=8$. Together with the two-level qubit, this gives a total optimization Hilbert-space dimension $N_h=2N_k=16$.

We seek any unitary evolution whose action on $\mathcal H_{\rm in}$ realizes the target isometry and whose corresponding output has no component outside $\mathcal H_{\rm out}$. We order the first four basis vectors as $\{\ket{g}_q\ket{0}_k,\ket{e}_q\ket{0}_k,\ket{g}_q\ket{1}_k,\ket{e}_q\ket{1}_k\}$, followed by the remaining states of the numerical truncation. The first two columns then span $\mathcal H_{\rm in}$, while the first four rows span $\mathcal H_{\rm out}$, and the propagated unitary can be partitioned as
\begin{equation}\label{utarg}
    U(\vec c,T)=
    \begin{pmatrix}
        V'(\vec c,T)&B_1'(\vec c,T)\\
        O'(\vec c,T)&B_2'(\vec c,T)
    \end{pmatrix}.
\end{equation}
The block dimensions are
\begin{equation}
\begin{aligned}
    V'&:\;4\times2,
    &\qquad O'&:\;(N_h-4)\times2,\\
    B_1'&:\;4\times(N_h-2),
    &\qquad B_2'&:\;(N_h-4)\times(N_h-2).
\end{aligned}
\end{equation}
Only the first block column is constrained by the isometry-synthesis task:
\begin{equation}
    V'(\vec c,T)=\hat V_{[k]},
    \qquad
    O'(\vec c,T)=0.
\end{equation}
The remaining blocks $B_1'$ and $B_2'$ are not specified, apart from the restrictions imposed by the unitarity of the full propagator.

Following Refs.~\cite{zyryd,zycq}, we quantify the target-isometry overlap and the final leakage by
\begin{align}
    F_V(\vec c)
    &=\operatorname{Tr}\!\left[
      \hat V_{[k]}^\dagger V'(\vec c,T)
      \right],\\
    F_O(\vec c)
    &=\operatorname{Tr}\!\left[
      O'^\dagger(\vec c,T)O'(\vec c,T)
      \right].
\end{align}
The control parameters are obtained by minimizing
\begin{equation}\label{eq:qoc_cost}
    \mathcal C(\vec c)
    =1-\frac{1}{D}\left|F_V(\vec c)\right|
    +\lambda F_O(\vec c),
    \qquad D=2,
\end{equation}
where $|F_V|/D$ is the normalized isometry overlap and $\lambda>0$ is the weight of the soft leakage penalty. This quantity is an optimization overlap and should not be identified with the process fidelity reconstructed from experimental process tomography.

To obtain gradients without finite-differencing the control parameters, we propagate the GOAT sensitivity equations together with the Schr\"odinger equation. For every parameter $c_k$,
\begin{equation}\label{eq:goat_sensitivity}
    \partial_t
    \begin{pmatrix}
        U\\
        \partial_{c_k}U
    \end{pmatrix}
    =-i
    \begin{pmatrix}
        H&0\\
        \partial_{c_k}H&H
    \end{pmatrix}
    \begin{pmatrix}
        U\\
        \partial_{c_k}U
    \end{pmatrix},
\end{equation}
where $H=H_{\rm int}/\hbar$ in the units used for numerical propagation. The blocks $V'$, $O'$ and their derivatives are extracted from $U(T)$ and $\partial_{c_k}U(T)$, and are then used to evaluate $\mathcal C$ and its analytic sensitivity gradient, up to the numerical accuracy of the ODE integration. This continuous-time construction does not require a piecewise-constant control ansatz.

The two drive quadratures are parameterized independently by analytic Fourier series. Denoting $\mu\in\{x,y\}$,
\begin{equation}
    f_\mu(t)=\sum_{j=1}^{j_{\max}}
    A_{\mu j}\sin(j\omega_\mu t).
\end{equation}
As in Refs.~\cite{zyryd,zycq}, the raw Fourier functions are passed through a smooth amplitude-bounding function and a turn-off envelope,
\begin{align}
    S_1[f_\mu(t)]
    &=b\left[
      1-\frac{2}{1+\exp\!\left(g_1f_\mu(t)/b\right)}
      \right],\\
    S_2(t,T)
    &=1-\frac{2}{1+\exp[g_2(T-t)]},
\end{align}
so that
\begin{equation}
    \Omega_\mu(t)=S_1[f_\mu(t)]S_2(t,T),
    \qquad \mu=x,y.
\end{equation}
Here $b$ sets the saturation amplitude of each quadrature, $g_1$ controls the steepness of the amplitude-bounding function, and $g_2$ sets the turn-off rate of the final envelope. Because the sine series vanishes at $t=0$ and $S_2(T,T)=0$, the resulting controls start and end at zero. The Fourier amplitudes $A_{\mu j}$ and frequencies $\omega_\mu$ constitute the optimization parameters $\vec c$; the amplitude bounds and smoothing constants are fixed pulse-parameterization settings.

We supply the analytic gradient to MATLAB's \texttt{fminunc} routine and use its quasi-Newton algorithm with a BFGS Hessian update. The Fourier coefficients and quadrature frequencies are initialized randomly. The optimized pulse implementing $\hat V_{\rm GHZ}$ is shown in Fig.~\ref{fig1}d of the main text.

\subsubsection{Advantage of isometry synthesis over unitary synthesis}

The essential distinction between the present optimization and fully specified unitary synthesis is the number of constrained input columns. A specified unitary fixes the evolution of an entire basis, whereas the isometry fixes only the $D=2$ columns associated with the physically populated input subspace $\mathcal H_{\rm in}$. The other $N_h-D$ columns, represented by $B_1'$ and $B_2'$ in Eq.~\eqref{utarg}, remain free. This enlarges the manifold of acceptable unitary extensions and gives the optimizer additional freedom to find controls that can be shorter or easier to realize than pulses for a particular fully specified unitary, such as a chosen CNOT extension~\cite{zyryd,zycq}.

\clearpage
\newpage

\section{Quantum process tomography}\label{sec:QPT}

The dynamics of a quantum system can be described by a completely positive (CP) linear map
\begin{equation}
    \mathcal{E}(\rho) = \sum_r A_r \rho A_r^\dagger \;,
\end{equation}
where $\rho$ is the initial state of the system. The condition $\sum_r A_r^\dagger A_r \leq \mathbb{1}$ ensures that $\text{Tr}[\mathcal{E}(\rho)] \leq \text{Tr}[\rho]$, such that $\mathcal{E}$ is trace-nonincreasing. If $\sum_r A_r^\dagger A_r = \mathbb{1}$, then $\mathcal{E}$ is trace-preserving (TP). A completely positive and trace-preserving map is referred to as a quantum channel.

Consider an input Hilbert space of dimension $d_{\mathrm{in}}$ and an output Hilbert space of dimension $d_{\mathrm{out}}$, together with Hermitian operator bases $\{E_m^{\mathrm{in}}\}_{m=0}^{d_{\mathrm{in}}^2-1}$ and $\{E_n^{\mathrm{out}}\}_{n=0}^{d_{\mathrm{out}}^2-1}$ satisfying
\begin{equation}
    \text{Tr}\!\left[(E_m^{\mathrm{in}})^\dagger E_{m'}^{\mathrm{in}}\right]=d_{\mathrm{in}}\delta_{mm'} \;,\qquad \text{Tr}\!\left[(E_n^{\mathrm{out}})^\dagger E_{n'}^{\mathrm{out}}\right]=d_{\mathrm{out}}\delta_{nn'} \;.
\end{equation}
For example, for a qubit one may choose the three Pauli matrices together with the $2\times2$ identity matrix. Expanding an input state as $\rho=\sum_m\rho_mE_m^{\mathrm{in}}$, the linearity of $\mathcal{E}$ allows the output state to be written as
\begin{equation}
    \mathcal{E}(\rho)=\sum_n\left(\sum_m\chi_n^m\rho_m\right)E_n^{\mathrm{out}} \;,
\end{equation}
where $\bm{\chi}$ is the matrix representation of the process $\mathcal{E}$ in the chosen input and output operator bases. It is a $d_{\mathrm{out}}^2\times d_{\mathrm{in}}^2$ matrix. For a Hermiticity-preserving map and Hermitian operator bases, the coefficients $\chi_n^m$ are real.

The central idea behind process tomography (PT) is to prepare $d_{\mathrm{in}}^2$ linearly independent input states $\{\rho_k\}_{k=0}^{d_{\mathrm{in}}^2-1}$, which span the operator space of the input system, and measure each output $\mathcal{E}(\rho_k)$ by standard quantum state tomography. Since $\mathcal{E}$ is a linear map, its matrix representation $\bm{\chi}$ can then be reconstructed from this set of measured input-output pairs.

We now specialize this framework to the process realized in our experiment, which maps an input state of the qubit to an output state of the joint qubit-phonon system,
\begin{equation}
    \mathcal{E}:\rho_Q^{\mathrm{in}}\longrightarrow\rho_{QP}^{\mathrm{out}} \;.
\end{equation}
Since our experiment deals with controlling the two lowest Fock states of the phonon mode, we describe it as a two-level system throughout, although the approach below generalizes straightforwardly to multilevel scenarios. We therefore have $d_{\mathrm{in}}=2$ and $d_{\mathrm{out}}=4$, such that $\bm{\chi}$ is a $16\times4$ matrix mapping the four input-qubit Pauli coefficients onto the sixteen output qubit-phonon Pauli coefficients.

Writing the elements of $\bm{\chi}$ as $\chi_{\alpha\beta}^{\gamma}$, where $\gamma\in\{0,x,y,z\}$ labels the input-qubit Pauli operator and $\alpha,\beta\in\{0,x,y,z\}$ label the output qubit and phonon Pauli operators, respectively (with $\sigma_0\equiv\mathbb{1}$), an input state
\begin{equation}
    \rho=\sum_\gamma\rho_\gamma\sigma_\gamma \;,\qquad \rho_\gamma=\tfrac{1}{2}\text{Tr}[\rho\,\sigma_\gamma] \;,
\end{equation}
is mapped as
\begin{equation}\label{supp:map}
    \mathcal{E}(\rho)=\sum_{\alpha,\beta}r_{\alpha\beta}\,\sigma_\alpha\otimes\sigma_\beta \;,\qquad
    r_{\alpha\beta}=\tfrac{1}{4}\text{Tr}\!\left[\mathcal{E}(\rho)\,\sigma_\alpha\otimes\sigma_\beta\right]=\sum_\gamma\chi_{\alpha\beta}^{\gamma}\rho_\gamma \;.
\end{equation}
We identify the pair $(\alpha,\beta)$ with the row index of $\bm{\chi}$ and $\gamma$ with its column index.

In order to experimentally reconstruct $\bm{\chi}$, it would be sufficient to prepare four linearly independent qubit states $\{\rho_Q^{\mathrm{in},i}\}_{i=1}^4$ and measure the corresponding output states $\{\rho_{QP}^{\mathrm{out},i}\}_{i=1}^4$. For each input state, we collect its four Pauli coefficients into the vector $\bm{r}_Q^{\mathrm{in},i}$, and similarly collect the sixteen two-qubit Pauli coefficients $r_{\alpha\beta}$ of the corresponding output state into $\bm{r}_{QP}^{\mathrm{out},i}$. Constructing
\begin{equation}
    R^{\mathrm{in}}=\left[\bm{r}_Q^{\mathrm{in},1},\ldots,\bm{r}_Q^{\mathrm{in},4}\right] \;,\qquad R^{\mathrm{out}}=\left[\bm{r}_{QP}^{\mathrm{out},1},\ldots,\bm{r}_{QP}^{\mathrm{out},4}\right] \;,
\end{equation}
the linear relation $R^{\mathrm{out}}=\bm{\chi}R^{\mathrm{in}}$ gives
\begin{equation}
    \bm{\chi}=R^{\mathrm{out}}(R^{\mathrm{in}})^{-1} \;.
\end{equation}
Since the Pauli coefficients of Hermitian operators are real, $R^{\mathrm{in}}$ and $R^{\mathrm{out}}$ are real matrices, which is why ordinary rather than conjugate transposes appear below.
This minimal approach, however, is sensitive to experimental imperfections, since it relies on only four noisy input-output measurements. We therefore reconstruct $\bm{\chi}$ from an overcomplete set by preparing the qubit in the six cardinal points of the Bloch sphere, $\rho_Q^{\mathrm{in},i}=\ket{\psi_i}\bra{\psi_i}$ with
\begin{equation}
    \{\ket{\psi_i}\}_{i=1}^6=\left\lbrace \ket{g},\ket{e}, \dfrac{\ket{g}+\ket{e}}{\sqrt{2}}, \dfrac{\ket{g}-\ket{e}}{\sqrt{2}}, \dfrac{\ket{g}+i\ket{e}}{\sqrt{2}}, \dfrac{\ket{g}-i\ket{e}}{\sqrt{2}} \right\rbrace \;,
\end{equation}
and measuring the corresponding six output density matrices $\{\rho_{QP}^{\mathrm{out},i}\}_{i=1}^6$. The matrices $R^{\mathrm{in}}$ and $R^{\mathrm{out}}$ are then of dimension $4\times6$ and $16\times6$, respectively. The resulting system of equations for $\bm{\chi}$ is overdetermined, and we obtain a least-squares estimate
\begin{equation}
    \bm{\chi}_{\rm LS}\equiv\underset{\bm{\chi}}{\text{argmin}}\,\left\|R^{\mathrm{out}}-\bm{\chi}R^{\mathrm{in}}\right\|_F^2 \;,
\end{equation}
where
\begin{equation}
    \|A\|_F=\sqrt{\sum_{ij}|A_{ij}|^2}
\end{equation}
is the Frobenius norm. Since the six cardinal states span the qubit operator space, $R^{\mathrm{in}}$ has full row rank, such that $R^{\mathrm{in}}(R^{\mathrm{in}})^T$ is invertible and the analytical least-squares solution is
\begin{equation}
    \bm{\chi}_{\rm LS}=R^{\mathrm{out}}(R^{\mathrm{in}})^T\left[R^{\mathrm{in}}(R^{\mathrm{in}})^T\right]^{-1} \;.
\end{equation}

The least-squares estimate is not guaranteed to correspond to a physical quantum channel. To impose complete positivity and trace preservation, we introduce the Choi matrix associated with $\bm{\chi}$,
\begin{equation}
    C^\chi=\sum_{i,j=0}^{1}e_{ij}\otimes\mathcal{E}(e_{ij}) \;,
\end{equation}
where $e_{ij}=\ket{i}\bra{j}$ acts on the input-qubit Hilbert space. Using Eq.~\eqref{supp:map}, this can be written as
\begin{equation}
    C^\chi=\sum_{i,j=0}^{1} e_{ij}\otimes \left[ \sum_{\alpha,\beta,\gamma} \chi_{\alpha\beta}^{\gamma} \Lambda_\gamma(e_{ij}) \sigma_\alpha\otimes\sigma_\beta \right] \;,
\end{equation}
where $\Lambda_\gamma(e_{ij})=\frac{1}{2}\text{Tr}[e_{ij}\sigma_\gamma]$ is the coefficient of $e_{ij}$ in the Pauli basis. Although $e_{ij}$ is not Hermitian for $i\neq j$, Eq.~\eqref{supp:map} extends to these operators by linearity. Since the Choi matrix acts on the tensor-product space of the input qubit and the output qubit-phonon system, its dimension is $(2\times4)\times(2\times4)=8\times8$. In the tensor products below, we use the ordering input qubit $\otimes$ output qubit $\otimes$ output phonon.

Choi's theorem states that $\mathcal{E}$ is completely positive if and only if the associated Choi matrix is positive semidefinite,
\begin{equation}
    C^\chi\succeq0 \;.
\end{equation}
Trace preservation requires $\text{Tr}[\mathcal{E}(\rho)]=\text{Tr}[\rho]$ for every input operator $\rho$. From Eq.~\eqref{supp:map}, this condition gives $\chi_{00}^{0}=\frac{1}{2}$, $\chi_{00}^{\gamma}=0$, for $\gamma\neq0$. Equivalently, the Choi matrix must satisfy $\text{Tr}_{\mathrm{out}}[C^\chi]=\mathbb{1}_{\mathrm{in}}$.
In the Pauli basis and with the tensor ordering specified above, this is equivalent to
\begin{equation}\label{supp:Cconstr}
    \text{Tr}\!\left[ C^\chi\, \sigma_0\otimes\sigma_0\otimes\sigma_0 \right]=2 \;,\qquad \text{Tr}\!\left[ C^\chi\, \sigma_\gamma\otimes\sigma_0\otimes\sigma_0 \right]=0 \quad\text{for}\quad\gamma\neq0 \;.
\end{equation}

To obtain a physical process from the experimentally estimated $\bm{\chi}_{\rm LS}$, we first construct the corresponding Choi matrix $C^{\chi_{\rm LS}}$ and then use a semidefinite program, implemented in Python with CVXPY, to find the Choi matrix $C^\chi$ minimizing
\begin{equation}
    \left\|C^\chi-C^{\chi_{\rm LS}}\right\|_F
\end{equation}
subject to the trace-preservation constraints of Eq.~\eqref{supp:Cconstr} and the complete-positivity constraint $C^\chi\succeq0$. The corresponding physical process matrix $\bm{\chi}$ is then obtained from $C^\chi$ through the inverse of the linear transformation defined above.

The process fidelity between the reconstructed and ideal processes is computed from their Choi matrices. Using the square-root fidelity convention employed throughout this work, we define
\begin{equation}
    F=\frac{ \text{Tr}\!\left[ \sqrt{\sqrt{C^\chi}\, C^\chi_{\mathrm{ideal}}\, \sqrt{C^\chi}} \right] }{ \sqrt{\text{Tr}[C^\chi]\, \text{Tr}[C^\chi_{\mathrm{ideal}}]} } \;.
\end{equation}

\clearpage
\newpage

\section{Witnessing genuine multipartite entanglement}\label{sec:GME}

An $N$-partite state $\sigma$ is called biseparable if it can be written as a convex mixture of states that are each separable with respect to some bipartition of the $N$ parties,
\begin{equation}
\sigma = \sum_k p_k \, \sigma^{(k)}_{A_k} \otimes \sigma^{(k)}_{B_k},
\end{equation}
where $p_k\geq0$, $\sum_k p_k=1$, $A_k \cup B_k = \{1,\dots,N\}$, $A_k \cap B_k = \varnothing$, and different terms may correspond to different bipartitions. A state that is not biseparable is called genuinely multipartite entangled (GME). Note that GME is strictly stronger than entanglement across every bipartition, since a mixture of states that are each separable across a different bipartition can be entangled across all bipartitions while still being biseparable~\cite{guhne_entanglement_2009}.

A Hermitian operator $\hat{\mathcal{W}}$ is a GME witness if
\begin{equation}
\mathrm{Tr}[\hat{\mathcal{W}}\sigma] \geq 0 \quad \text{for all biseparable states } \sigma,
\end{equation}
so that observing $\mathrm{Tr}[\hat{\mathcal{W}}\rho] < 0$ for a state $\rho$ certifies that $\rho$ is GME. Since the biseparable states form a convex set whose extreme points are pure states $\ket{\vect{a}}_A\otimes\ket{\vect{b}}_B$ that are product with respect to some bipartition $A|B$, it suffices to verify this condition on such pure product states.

Throughout this section we denote by
\begin{equation}
\mathcal{F}^2_\psi = \bra{\psi}\rho\ket{\psi}
\end{equation}
the squared fidelity of $\rho$ with a pure target state $\ket{\psi}$, where $\mathcal{F}_\psi$ is the fidelity defined in the main text.
All witnesses below act on the computational subspace $\{\ket{0}_k,\ket{1}_k\}$ of each phonon mode $k$, in which the density matrices are reconstructed.

\subsection{Fidelity-based witnesses}

For a target state $\ket{\psi}$, the operator
\begin{equation}\label{eq:SM_fidwit}
\hat{\mathcal{W}}_\psi = c_\psi\,\mathbb{1} - \ket{\psi}\bra{\psi},
\qquad
c_\psi = \max_{\sigma\in\mathrm{bisep}} \bra{\psi}\sigma\ket{\psi},
\end{equation}
is a GME witness by construction, and $\mathrm{Tr}[\hat{\mathcal{W}}_\psi\rho] = c_\psi - \mathcal{F}_\psi^2$~\cite{bourennane_experimental_2004}. Observing $\mathcal{F}_\psi^2 > c_\psi$, i.e.\ $\mathcal{F}_\psi>\sqrt{c_\psi}$, therefore certifies GME.
The constant $c_\psi$ is the maximal overlap of $\ket{\psi}$ with a pure product state across some bipartition, and is given by the square of the largest Schmidt coefficient of $\ket{\psi}$, maximized over all bipartitions~\cite{bourennane_experimental_2004}.

\textbf{Graph states.} For a connected graph state $\ket{G_N}$, the reduced state on either side of any bipartition is maximally mixed on a support of dimension $2^r$, where $r\geq1$ because at least one edge of a connected graph crosses the cut. All nonzero Schmidt coefficients are therefore equal to $2^{-r/2}$, and bipartitions with $r=1$ always exist (e.g.\ separating a single vertex or, for GHZ states, any cut), so that $c_{G}=1/2$~\cite{TothPRL05}. This yields the witness
\begin{equation}
\hat{\mathcal{W}}_3(N) = \tfrac{1}{2}\mathbb{1} - \ket{G_N}\bra{G_N},
\end{equation}
which detects GME whenever $\mathcal{F}^2_{G}>1/2$, i.e.\ $\mathcal{F}_G>1/\sqrt{2}$. Both the $N$-partite GHZ state and the linear cluster state are, up to local unitaries, connected graph states (the GHZ state corresponds to the star graph).
Since local unitaries leave the set of biseparable states invariant, the bound $c=1/2$ holds for the target states as prepared in our experiment. For $N=2$, both reduce to a Bell state.

\textbf{$W$ states.} Across a bipartition in which $K$ modes are separated from the remaining $N-K$, the $W$ state has the Schmidt decomposition
\begin{equation}
\ket{W_N} = \sqrt{\tfrac{K}{N}}\,\ket{W_K}\ket{0\cdots0} + \sqrt{\tfrac{N-K}{N}}\,\ket{0\cdots0}\ket{W_{N-K}},
\end{equation}
so that $c_{W}=\max_K \max(K,N-K)/N=(N-1)/N$, attained for $K=1$ by the states $\ket{0}_k\ket{W_{N-1}}$~\cite{bourennane_experimental_2004,haffner_scalable_2005}. The corresponding witness
\begin{equation}
\hat{\mathcal{W}}_1(N) = \frac{N-1}{N}\mathbb{1} - \ket{W_N}\bra{W_N}
\end{equation}
detects GME whenever $\mathcal{F}^2_{W}>(N-1)/N$, i.e.\ $\mathcal{F}_W>\sqrt{(N-1)/N}$. The large overlap of $\ket{W_N}$ with biseparable states makes this threshold considerably more demanding than for graph states, and it approaches unity as $N$ grows. Since the bound is attained, $\hat{\mathcal{W}}_1(N)$ is optimal among witnesses of the form $c\,\mathbb{1}-\ket{W_N}\bra{W_N}$; improving on it requires using more information about the state than its fidelity alone.

\subsection{Tailored witness for $W$ states}

Following Ref.~\cite{haffner_scalable_2005}, we consider the $N$ biseparable states
\begin{equation}
\ket{\mathrm{BS}_k} = \ket{0}_k \otimes \ket{W_{N-1}}, \qquad k=1,\dots,N,
\end{equation}
in which mode $k$ carries no excitation and a single excitation is delocalized over the remaining $N-1$ modes. These are precisely the biseparable states attaining the maximal overlap $(N-1)/N$ with $\ket{W_N}$. We define the operator
\begin{equation}\label{eq:SM_Q}
\hat{Q}(N,\beta) = \ket{W_N}\bra{W_N} - \beta \sum_{k=1}^{N} \ket{\mathrm{BS}_k}\bra{\mathrm{BS}_k},
\end{equation}
with $\beta\geq0$. The first term rewards overlap with the target state, while the second penalizes overlap with the biseparable states $\ket{\mathrm{BS}_k}$. Compared to Ref.~\cite{haffner_scalable_2005}, we have set the prefactor of the first term to one without loss of generality, since rescaling $\hat{Q}$ by a positive constant rescales the resulting witness by the same constant; our $\beta$ thus corresponds to the ratio $\beta/\alpha$ in Ref.~\cite{haffner_scalable_2005}.

The biseparable bound
\begin{equation}
\gamma(N,\beta) = \max_{\sigma\in\mathrm{bisep}} \mathrm{Tr}[\hat{Q}(N,\beta)\,\sigma]
= \max_{A|B}\;\max_{\ket{\vect{a}},\ket{\vect{b}}} \bra{\vect{a}}\bra{\vect{b}}\hat{Q}(N,\beta)\ket{\vect{a}}\ket{\vect{b}}
\end{equation}
can be computed efficiently as follows. Since $\hat{Q}$ is supported on the single-excitation subspace, $\bra{\psi}\hat{Q}\ket{\psi}$ depends only on the single-excitation amplitudes $v_j=\bracket{1_j}{\psi}$, where $\ket{1_j}$ denotes one excitation in mode $j$. Using $\bracket{W_N}{\psi}=S/\sqrt{N}$ and $\bracket{\mathrm{BS}_k}{\psi}=(S-v_k)/\sqrt{N-1}$, with $S=\sum_j v_j$, one finds
\begin{equation}\label{eq:SM_Qexp}
\bra{\psi}\hat{Q}(N,\beta)\ket{\psi} = \abs{S}^2\left[\frac{1}{N}-\beta\,\frac{N-2}{N-1}\right] - \frac{\beta}{N-1}\sum_{j}\abs{v_j}^2 .
\end{equation}
For a product state $\ket{\vect{a}}_A\ket{\vect{b}}_B$, the single-excitation amplitudes are $v_j=a_0 b_j$ for $j\in B$ and $v_j=b_0a_j$ for $j\in A$, where $a_0$ ($b_0$) is the vacuum amplitude and $a_j$ ($b_j$) the single-excitation amplitudes of $\ket{\vect{a}}$ ($\ket{\vect{b}}$). Components with two or more excitations within $A$ or $B$ do not contribute to Eq.~\eqref{eq:SM_Qexp} and only reduce the available normalization, so they can be set to zero. Moreover, for a fixed norm of the single-excitation part within each partition, $\abs{S}$ is maximized and $\sum_j\abs{v_j}^2$ is left unchanged by choosing all amplitudes within a partition equal in magnitude and phase (Cauchy--Schwarz inequality), and the relative phase between the two partitions can be chosen such that the two contributions to $S$ add constructively. Whenever the bracket in Eq.~\eqref{eq:SM_Qexp} is negative, $\gamma=0$, which is attained by the vacuum state. The optimization therefore reduces to
\begin{equation}
\ket{\vect{a}} = \cos\theta_A \ket{0\cdots0} + \sin\theta_A \ket{W_K}, \qquad
\ket{\vect{b}} = \cos\theta_B \ket{0\cdots0} + \sin\theta_B \ket{W_{N-K}},
\end{equation}
i.e.\ to a maximization over two angles for each bipartition size $K=1,\dots,\lfloor N/2\rfloor$, equivalent to the four-parameter optimization with two normalization constraints of Ref.~\cite{haffner_scalable_2005}. For example, we obtain $\gamma(3,0.298)=0.2260$ and $\gamma(4,0.287)=0.0832$, reproducing the values of Ref.~\cite{haffner_scalable_2005} after rescaling, and $\gamma(4,0.300)=0.0577$.

The GME witness is then defined as
\begin{equation}\label{eq:SM_W2}
\hat{\mathcal{W}}_2(N,\beta) = \gamma(N,\beta)\,\mathbb{1}_{(2)} - \hat{Q}(N,\beta),
\end{equation}
where $\mathbb{1}_{(2)}$ is the projector onto the subspace with at most two excitations in total~\cite{haffner_scalable_2005}. To see that $\hat{\mathcal{W}}_2$ is positive on all biseparable states, let $\ket{\vect{a}'}$ and $\ket{\vect{b}'}$ be the (unnormalized) projections of $\ket{\vect{a}}$ and $\ket{\vect{b}}$ onto their vacuum and single-excitation components. The state $\ket{\vect{a}'}\ket{\vect{b}'}$ has the same single-excitation amplitudes as $\ket{\vect{a}}\ket{\vect{b}}$, contains at most two excitations, and is orthogonal to the remaining components of $\ket{\vect{a}}\ket{\vect{b}}$. Hence
\begin{equation}
\bra{\vect{a}}\bra{\vect{b}}\hat{Q}\ket{\vect{a}}\ket{\vect{b}}
= \bra{\vect{a}'}\bra{\vect{b}'}\hat{Q}\ket{\vect{a}'}\ket{\vect{b}'}
\leq \gamma\,\big\|\ket{\vect{a}'}\ket{\vect{b}'}\big\|^2
\leq \gamma\,\bra{\vect{a}}\bra{\vect{b}}\mathbb{1}_{(2)}\ket{\vect{a}}\ket{\vect{b}},
\end{equation}
where we used $\gamma\geq0$. By convexity, $\mathrm{Tr}[\hat{\mathcal{W}}_2\sigma]\geq0$ for every biseparable $\sigma$. Restricting the identity to at most two excitations is thus the most restrictive choice of this form that preserves validity: a projector onto at most one excitation would not suffice, since $\ket{\vect{a}'}\ket{\vect{b}'}$ generally contains a two-excitation component. Since $\mathrm{Tr}[\mathbb{1}_{(2)}\rho]\leq 1$, $\hat{\mathcal{W}}_2(N,0)$ is at least as sensitive as $\hat{\mathcal{W}}_1(N)$.

Since $\hat{\mathcal{W}}_2(N,\beta)$ is a valid GME witness for every $\beta\geq0$, $\beta$ can be chosen to minimize $\mathrm{Tr}[\hat{\mathcal{W}}_2(N,\beta)\rho]$ on the experimental state. For the ideal state, $\bra{W_N}\hat{Q}\ket{W_N}=1-\beta(N-1)$, and the most negative value is obtained for $\beta=0$. 
Nonzero $\beta$ becomes advantageous for imperfect states whose overlap with the states $\ket{\mathrm{BS}_k}$ is reduced relative to their overlap with $\ket{W_N}$.

\subsection{Local-filtering-optimized witnesses}\label{sec:filtering}

The detection capability of any of the above witnesses can be improved by local filtering~\cite{haffner_scalable_2005,kiesel_experimental_2007}. Let
\begin{equation}
\hat F=\hat A_1\otimes\cdots\otimes\hat A_N,
\end{equation}
where each $\hat A_k$ is an arbitrary complex $2\times2$ matrix acting on the computational subspace of mode $k$, and let $\hat{\mathcal W}$ be any GME witness. The filtered operator
\begin{equation}
\hat{\mathcal W}'=\hat F^\dagger\hat{\mathcal W}\hat F
\end{equation}
is again a GME witness. Indeed, for every biseparable $\sigma=\sum_k p_k\,\sigma^{(k)}_{A_k}\otimes\sigma^{(k)}_{B_k}$, the operator
\begin{equation}
\hat F\sigma\hat F^\dagger=\sum_k p_k\,\Big(\hat F_{A_k}\sigma^{(k)}_{A_k}\hat F_{A_k}^\dagger\Big)\otimes\Big(\hat F_{B_k}\sigma^{(k)}_{B_k}\hat F_{B_k}^\dagger\Big),
\end{equation}
with $\hat F_{A_k}=\bigotimes_{j\in A_k}\hat A_j$ and analogously for $B_k$, is positive and biseparable up to normalization, so that
\begin{equation}
\mathrm{Tr}\big[\hat{\mathcal W}'\sigma\big]=\mathrm{Tr}\big[\hat{\mathcal W}\hat F\sigma\hat F^\dagger\big]\geq0.
\end{equation}
The only requirement is that $\hat F$ has product form with respect to the $N$ modes; the $\hat A_k$ need be neither unitary nor invertible.
Equivalently, $\mathrm{Tr}[\hat{\mathcal W}'\rho]<0$ states that the filtered state $\rho_F=\hat F\rho\hat F^\dagger/\mathrm{Tr}[\hat F\rho\hat F^\dagger]$, which can be obtained from $\rho$ with nonzero probability by local operations, is GME. Since local operations cannot create GME, $\rho$ itself must be GME.
Because rescaling $\hat F$ by a positive constant rescales $\mathrm{Tr}[\hat{\mathcal W}'\rho]$ and its statistical uncertainty by the same factor, we fix the normalization by $\mathrm{Tr}[\hat F\rho\hat F^\dagger]=1$, so that $\mathrm{Tr}[\hat{\mathcal W}'\rho]=\mathrm{Tr}[\hat{\mathcal W}\rho_F]$, and assess the significance of a violation through its ratio to the statistical uncertainty, which is independent of this choice.
Local filtering can compensate for specific imperfections of the prepared state, such as unequal populations of the branches of the superposition or residual local phases. In particular, the local virtual phase rotations used in the main text to compensate the relative phases of the target state are a special, unitary, case of local filtering.

The same argument justifies the restriction to the computational subspace of each phonon mode. The projector $\hat P=\bigotimes_k(\ket{0}_k\bra{0}+\ket{1}_k\bra{1})$ is itself a local filter, so that GME of the (renormalized) projected state implies GME of the full multimode state, provided that the tomographic reconstruction faithfully represents this projected state.

If all $\hat A_k$ are invertible, the filtering is reversible by the local filter $\hat F^{-1}$. For pure states, invertible local operators are precisely the transformations relating states within the same class under stochastic local operations and classical communication (SLOCC)~\cite{dur_three_2000}, and in Ref.~\cite{kiesel_experimental_2007} filtered witnesses were used to discriminate between SLOCC classes. We emphasize that here we use the filtered witnesses only to certify GME, and not membership in a particular SLOCC class.

\clearpage
\newpage

\section{Definition of multimode $W$, GHZ and cluster states}
\label{sec:multimode_states}

We consider a system of $N$ bosonic modes, each described by annihilation and creation operators
$\hat a_k$ and $\hat a_k^\dagger$ $(k=1,\ldots,N)$ satisfying the canonical commutation relations
\begin{equation}
[\hat a_k,\hat a_{k'}^\dagger]=\delta_{k,k'},\qquad
[\hat a_k,\hat a_{k'}]=[\hat a_k^\dagger,\hat a_{k'}^\dagger]=0.
\end{equation}
Acting on the vacuum state $\ket{0\dots0}$, the creation operators generate Fock states according to
\begin{equation}
\hat a_k^\dagger \ket{n_k}_k = \sqrt{n_k+1}\,\ket{n_k+1}_k,
\end{equation}
where $\ket{n_k}_k$ denotes the Fock state with $n_k$ excitations in mode $k$, and $\hat n_k\equiv\hat a_k^\dagger\hat a_k$ is the corresponding number operator.
The multimode Fock basis is given by
\begin{equation}
\ket{\mathbf{n}}=\ket{n_1 n_2\ldots n_N}
\equiv \prod_{k=1}^N \frac{(\hat a_k^\dagger)^{n_k}}{\sqrt{n_k!}}\,\ket{0\dots0}.
\end{equation}

In the following, we restrict attention to states defined within the lowest two Fock levels
$n_k\in\{0,1\}$ for each mode, which allows a direct correspondence with $N$-qubit states.
Within this subspace, we introduce three important classes of genuinely multipartite entangled states: $W$ states, GHZ states, and cluster states. For $N=2$, all three reduce, up to local unitaries, to a Bell state.

\subsection{$W$ states}

The $N$-partite $W$ state is defined as
\begin{equation}
|W_N\rangle
= \frac{1}{\sqrt{N}}\sum_{k=1}^N \hat a_k^\dagger \ket{0\dots0} = \frac{1}{\sqrt{N}}\sum_{k=1}^N \ket{1_k},
\qquad
\ket{1_k} \equiv \hat a_k^\dagger\ket{0\dots0},
\label{eq:WN_def}
\end{equation}
i.e.\ a uniform superposition of all computational basis states containing exactly one excitation.

The $W$ state belongs entirely to the single-excitation subspace of the total number operator
\begin{equation}
\hat N_{\mathrm{tot}}=\sum_{k=1}^N \hat n_k,
\end{equation}
and satisfies
\begin{equation}
\hat N_{\mathrm{tot}}|W_N\rangle = |W_N\rangle.
\end{equation}
As a consequence, all components of $|W_N\rangle$ share the same total excitation number.
As we will see, this property underlies the robustness of $W$ states against collective phase noise generated by $\hat N_{\mathrm{tot}}$.

$W$ states are genuinely multipartite entangled for all $N\geq 2$ and exhibit the characteristic feature that, for $N\geq3$, entanglement persists after tracing out one subsystem: the reduced state of the remaining $N-1$ modes, $\frac{N-1}{N}|W_{N-1}\rangle\langle W_{N-1}| + \frac{1}{N}\ket{0\dots0}\bra{0\dots0}$, with $\ket{0\dots0}$ here the vacuum of the remaining $N-1$ modes, remains genuinely multipartite entangled~\cite{dur_three_2000}.
This makes $W$ states particularly robust against the loss of individual modes, in the sense of tracing them out, which is distinct from the local energy loss treated in Sec.~\ref{sec:local_decoherence}.

\subsection{GHZ states}

The $N$-partite Greenberger--Horne--Zeilinger (GHZ) state is defined as
\begin{equation}
|\mathrm{GHZ}_N\rangle
= \frac{1}{\sqrt{2}}\Big(\ket{0\dots0} + \ket{1\dots1}\Big).
\label{eq:GHZN_def}
\end{equation}
This corresponds to a coherent superposition of the vacuum state and the state with one excitation in every mode. When these two states label matrix elements, we abbreviate them as $0^N$ and $1^N$, e.g.\ $\rho_{0^N,1^N}\equiv\bra{0\dots0}\rho\ket{1\dots1}$.

The two components of $|\mathrm{GHZ}_N\rangle$ belong to different eigenspaces of the total number operator:
\begin{equation}
\hat N_{\mathrm{tot}}\ket{0\dots0} = 0,\qquad
\hat N_{\mathrm{tot}}\ket{1\dots1} = N\,\ket{1\dots1}.
\end{equation}
Thus, $|\mathrm{GHZ}_N\rangle$ is not an eigenstate of $\hat N_{\mathrm{tot}}$ but a superposition of two sectors
with excitation-number difference $|\Delta n| = N$.

GHZ states exhibit maximal nonclassical correlations, achieving the largest possible violation of multipartite Bell-type (Mermin) inequalities.
However, they are extremely fragile: tracing out a single mode renders the remaining state separable.
Similarly, GHZ states are highly sensitive to collective phase noise: a fluctuating common phase $\phi$, imprinted on all modes by $e^{-i\phi \hat N_{\mathrm{tot}}}$, generates a relative phase $N\phi$ between the two branches, so that for zero-mean Gaussian-distributed $\phi$ with variance $\sigma_\phi^2$ the coherence decays as $\langle e^{iN\phi}\rangle = e^{-N^2\sigma_\phi^2/2}$, leading to a characteristic superdecoherence with a decay exponent scaling as $N^2$.

\subsection{Cluster states}

Cluster states form a distinct class of multipartite entangled states that play a central role
in measurement-based quantum computation.
They are most naturally defined as stabilizer states on a graph; for a general graph these are usually referred to as graph states, the term cluster state being commonly reserved for lattice graphs~\cite{hein_multiparty_2004}.

Consider a graph $G=(V,E)$ with vertices $V=\{1,\ldots,N\}$ corresponding to modes and edges $E$
specifying interactions.
The graph state associated with $G$ is defined (in the qubit subspace $n_k\in\{0,1\}$) as the
unique simultaneous $+1$ eigenstate of the stabilizer operators
\begin{equation}
\hat K_k = \hat\sigma_x^{(k)} \prod_{j\in \mathcal{N}(k)} \hat\sigma_z^{(j)},
\qquad k=1,\ldots,N,
\label{eq:cluster_stabilizers}
\end{equation}
where $\mathcal{N}(k)$ denotes the set of neighbors of vertex $k$, and $\hat\sigma_{x,z}^{(k)}$
act on the $\{\ket{0}_k,\ket{1}_k\}$ subspace of mode $k$.

Equivalently, a graph state can be constructed by preparing all modes in the superposition state
$|+\rangle = (|0\rangle+|1\rangle)/\sqrt{2}$ and applying controlled-phase (CZ) gates,
\begin{equation}
\mathrm{CZ}_{ij}\ket{n_i}_i\ket{n_j}_j = (-1)^{n_i n_j}\,\ket{n_i}_i\ket{n_j}_j,
\end{equation}
between all pairs of modes connected by an edge:
\begin{equation}
\ket{G_N}
= \prod_{\{i,j\}\in E} \mathrm{CZ}_{ij}\, |+\rangle^{\otimes N},
\end{equation}
which is the state denoted $\ket{G_N}$ in the main text.

Unlike $W$ states, which have a fixed total excitation number, and GHZ states, which span two excitation-number sectors, graph states are equal-magnitude superpositions of all $2^N$ computational basis states, and therefore contain all excitation numbers $0,\dots,N$.
Their entanglement structure is highly nontrivial and depends on the underlying graph.
Graph states are robust against local measurements: a $\hat\sigma_z$ measurement of a mode removes the corresponding vertex and its edges, projecting the remaining modes onto the graph state of the reduced graph, up to local Pauli corrections~\cite{BriegelPRL01,hein_multiparty_2004}, a key feature enabling measurement-based quantum computation.

\vspace{3mm}
A particularly important subclass of cluster states is given by \textbf{linear cluster states}, corresponding to a one-dimensional chain graph with vertices $V=\{1,\ldots,N\}$ and edges $E=\{\{1,2\},\{2,3\},\ldots,\{N-1,N\}\}$.
The associated stabilizer generators take the explicit form
\begin{align}
\hat K_1 &= \hat\sigma_x^{(1)}\hat\sigma_z^{(2)}, \nonumber\\
\hat K_k &= \hat\sigma_z^{(k-1)}\hat\sigma_x^{(k)}\hat\sigma_z^{(k+1)},
\qquad k=2,\ldots,N-1, \nonumber\\
\hat K_N &= \hat\sigma_z^{(N-1)}\hat\sigma_x^{(N)}.
\end{align}
The linear cluster state $|\mathrm{CL}_N^{\mathrm{lin}}\rangle$ is defined as the unique common
$+1$ eigenstate of all $\hat K_k$.

Equivalently, $|\mathrm{CL}_N^{\mathrm{lin}}\rangle$ can be prepared by initializing each mode in the superposition state $|+\rangle=(|0\rangle+|1\rangle)/\sqrt{2}$ and applying controlled-phase (CZ) gates sequentially between nearest-neighbor modes along the chain,
\begin{equation}
|\mathrm{CL}_N^{\mathrm{lin}}\rangle
= \prod_{k=1}^{N-1}\mathrm{CZ}_{k,k+1}\,|+\rangle^{\otimes N}.
\end{equation}
The sequential protocol described in the main text, with the qubit initialized in $\ket{+}_q=(\ket{g}_q+\ket{e}_q)/\sqrt2$, the isometry $\ket{g}_q\ket{0}_k\rightarrow\ket{+}_q\ket{0}_k$, $\ket{e}_q\ket{0}_k\rightarrow\ket{-}_q\ket{1}_k$, with $\ket{-}_q=(\ket{g}_q-\ket{e}_q)/\sqrt2$, applied to modes $1,\dots,N-1$, and a final iSWAP to mode $N$, prepares this state up to the local phase $\ket{1}_N\rightarrow-i\ket{1}_N$ on mode $N$, which is removed by the local virtual rotations described in the main text.

Linear cluster states implement a ``quantum wire'': sequential single-mode measurements allow a logical qubit to be transported along the chain while applying arbitrary single-qubit rotations, whereas universal measurement-based quantum computation requires cluster states on two-dimensional graphs~\cite{RaussendorfPRA03}.
Their entanglement structure is less symmetric than that of GHZ or $W$ states, but they exhibit high resilience under local measurements: measuring an interior mode in the $\hat\sigma_z$ basis removes that site together with its two edges, splitting the chain into two independent, shorter linear cluster states (up to local Pauli corrections depending on the measurement outcome), while measuring an endpoint mode instead yields a single linear cluster state on the remaining $N-1$ modes.
Propagating quantum information along the chain, as in a cluster-state quantum wire, instead relies on sequential measurements in the equatorial ($x$--$y$) plane of the Bloch sphere, starting from the end of the chain that carries the logical qubit; each measurement teleports the logical qubit to the next site, up to a known single-qubit unitary and outcome-dependent Pauli corrections~\cite{RaussendorfPRA03}.
This property makes linear cluster states an important building block for scalable architectures based on bosonic or phononic modes.

\clearpage
\newpage

\section{Decoherence of multimode states under local loss and dephasing}
\label{sec:local_decoherence}

For each mode $k$, local energy relaxation and pure dephasing are described by the jump operators
\begin{subequations}\label{eq:jumps}
\begin{align}
    \hat L_1^{(k)} & = \sqrt{\gamma_1^{(k)}}\,\hat a_k       && \text{(amplitude damping)},\\
    \hat L_\varphi^{(k)} & = \sqrt{2\gamma_{\varphi}^{(k)}}\;\hat n_k && \text{(pure dephasing)}. \label{eq:dephOp}
\end{align}
\end{subequations}
Here, $\gamma_1^{(k)}=1/T_{1,k}$ and $\gamma_\varphi^{(k)}=1/T^\ast_{2,k}-1/(2T_{1,k})$, with $T_{1,k}$ and $T^\ast_{2,k}$ the energy relaxation and Ramsey decoherence times of mode $k$ given in the main text.

The master equation for the density operator $\rho$ is
\begin{equation}\label{eq:master}
    \dot\rho = \sum_{k=1}^N\Bigl[\gamma_1^{(k)}\,\mathcal{D}[\hat a_k]\rho + 2\gamma_{\varphi}^{(k)}\,\mathcal{D}[\hat n_k]\rho \Bigr]
    \;,\quad\text{where} \qquad
    \mathcal{D}[\hat L]\rho = \hat L\rho \hat L^{\dagger} - \dfrac12\{\hat L^{\dagger}\hat L,\rho\}.
\end{equation}
Because each jump operator acts on a single mode, all channels commute and can be treated independently: the resulting dynamics factorizes as $\rho(t)=\left(\bigotimes_{k=1}^N\hat{\Phi}_k\right)[\rho(0)]$, where $\hat{\Phi}_k$ is the single-mode amplitude- and phase-damping channel with rates $\gamma_1^{(k)},\gamma_\varphi^{(k)}$.

Each mode is truncated to $\{\ket{0}_k,\ket{1}_k\}$, which is exact since Eq.~\eqref{eq:master} never populates higher Fock states.
We compute analytic fidelity decays for the target
states $\psi\in\{W_N,\mathrm{GHZ}_N,\mathrm{CL}^{\mathrm{lin}}_N\}$ and for general graph states, allowing all $\gamma_1^{(k)}$ and
$\gamma_\varphi^{(k)}$ to be distinct.

A useful rule implied by Eq.~\eqref{eq:master} is that, for two computational basis states
$\ket{\mathbf{m}}=\ket{m_1\ldots m_N}$ and $\ket{\mathbf{n}}=\ket{n_1\ldots n_N}$ with $m_k,n_k\in\{0,1\}$,
the coherence $\rho_{\mathbf{mn}}=\bra{\mathbf m}\rho\ket{\mathbf n}$ decays under pure dephasing as
\begin{equation}
\dot\rho_{\mathbf{mn}}\big|_{\varphi}
=-\left(\sum_{k: m_k\neq n_k}\gamma_\varphi^{(k)}\right)\rho_{\mathbf{mn}},
\label{eq:dephasing_rule}
\end{equation}
while amplitude damping contributes the decay rate $\tfrac12\sum_{k}\gamma_1^{(k)}(m_k+n_k)$.
This combined rule gives the exact decay rate whenever the equation of motion for $\rho_{\mathbf{mn}}$ is closed, i.e.\ receives no population feed-in from other matrix elements under the dynamics, as is the case for all single-excitation elements of $\ket{W_N}\bra{W_N}$ and for the coherence of $\ket{\mathrm{GHZ}_N}\bra{\mathrm{GHZ}_N}$ below. It is convenient to define, for two modes $j,k$,
\begin{equation}
\Gamma_{jk} \equiv \frac{\gamma_1^{(j)}+\gamma_1^{(k)}}{2}+\gamma_\varphi^{(j)}+\gamma_\varphi^{(k)},
\label{eq:Gamma_def}
\end{equation}
which will recur throughout this section. For states with a richer entanglement structure, such as the cluster state, several matrix elements are instead fed by population transfer from other coherences over time and no longer follow a single exponential; in that case the full closed form must be obtained from the explicit channel decomposition, as done in Sec.~\ref{sec:cl2} below.

\subsection{Notation}

Throughout this section, $\mathcal{F}^2_\psi(t)=\bra{\psi}\rho(t)\ket{\psi}$ denotes the square of the fidelity $\mathcal{F}$ defined in the main text, with respect to the target state $\ket{\psi}$, with $\rho(0)=\ket{\psi}\bra{\psi}$. We use the convention $\hat\sigma_z\ket{0}=\ket{0}$, $\hat\sigma_z\ket{1}=-\ket{1}$, and define for each mode $k$ the population and coherence decay factors
\begin{equation}\label{eq:p_r_def}
p_k(t)\equiv e^{-\gamma_1^{(k)}t},\qquad
r_k(t)\equiv e^{-\lambda_k t},\qquad
\lambda_k\equiv\frac{\gamma_1^{(k)}}{2}+\gamma_\varphi^{(k)},
\end{equation}
so that $\Gamma_{jk}=\lambda_j+\lambda_k$ in Eq.~\eqref{eq:Gamma_def} and $e^{-\Gamma_{jk}t}=r_jr_k$. For identical rates on all modes we drop the mode index, $p=e^{-\gamma_1t}$ and $r=e^{-(\gamma_1/2+\gamma_\varphi)t}$.
As in the main text, the normalized coherence is $\mathcal{C}(t)=\sum_{\mathbf m\neq\mathbf n}M_{\mathbf{mn}}\abs{\rho_{\mathbf{mn}}(t)}/\sum_{\mathbf m\neq\mathbf n}M_{\mathbf{mn}}\abs{\rho_{\mathbf{mn}}(0)}$, where $M_{\mathbf{mn}}=1$ if $\rho_{\mathbf{mn}}$ is non-zero in the ideal state and $M_{\mathbf{mn}}=0$ otherwise, and for $W$ states we also use the pairwise coherences $c_{kl}(t)=\abs{\rho_{1_k,1_l}(t)}/\abs{\rho_{1_k,1_l}(0)}$.

\subsection{$W$ states}

Under Eq.~\eqref{eq:master}, the single-excitation sector only decays into the vacuum, which is orthogonal to $\ket{W_N}$. The equations of motion for the single-excitation elements are therefore closed, and the rule of Eq.~\eqref{eq:dephasing_rule} and below gives $\rho_{1_k,1_k}(t)=p_k/N$ and $\rho_{1_k,1_l}(t)=r_kr_l/N$ for $k\neq l$. Hence
\begin{equation}
\mathcal{F}^2_{W_N}(t)=\frac{1}{N^2}\Big[\sum_{k=1}^N p_k+2\sum_{k<l}r_kr_l\Big],
\qquad
c_{kl}(t)=e^{-\Gamma_{kl}t},
\qquad
\mathcal{C}_{W_N}(t)=\binom{N}{2}^{-1}\sum_{k<l}e^{-\Gamma_{kl}t}.
\label{eq:F_WN}
\end{equation}
For $N=2$, this gives $\mathcal{F}^2_{W_2}=\tfrac14(p_1+p_2)+\tfrac12e^{-\Gamma_{12}t}$ and $\mathcal{C}_{W_2}=e^{-\Gamma_{12}t}$; for $N=3$, $\mathcal{F}^2_{W_3}=\tfrac19\big[\sum_kp_k+2\sum_{k<l}e^{-\Gamma_{kl}t}\big]$ and $\mathcal{C}_{W_3}=\tfrac13\sum_{k<l}e^{-\Gamma_{kl}t}$.
For identical rates,
\begin{equation}
\mathcal{F}^2_{W_N}(t)=\frac1N\Big[e^{-\gamma_1t}+(N-1)\,e^{-(\gamma_1+2\gamma_\varphi)t}\Big],
\qquad
\mathcal{C}_{W_N}(t)=e^{-(\gamma_1+2\gamma_\varphi)t}.
\label{eq:F_WN_sym}
\end{equation}
The coherence decay rate is thus independent of $N$, and for $N\to\infty$ the squared fidelity approaches the coherence decay $e^{-(\gamma_1+2\gamma_\varphi)t}$.

\subsection{GHZ states}

The coherence $\rho_{0^N,1^N}$ cannot be fed by any other element, and the populations follow from the independent decay of each excitation: $\rho_{1^N,1^N}=\tfrac12\prod_kp_k$, $\rho_{0^N,0^N}=\tfrac12+\tfrac12\prod_k(1-p_k)$, and $\rho_{0^N,1^N}=\tfrac12\prod_kr_k$. All other populations generated by the decay are orthogonal to the target, so that
\begin{equation}
\mathcal{F}^2_{\mathrm{GHZ}_N}(t)=\frac14\Big[1+\prod_{k=1}^N\big(1-p_k\big)+\prod_{k=1}^N p_k\Big]+\frac12\prod_{k=1}^N r_k,
\qquad
\mathcal{C}_{\mathrm{GHZ}_N}(t)=e^{-\sum_k\lambda_k t}.
\label{eq:F_GHZN}
\end{equation}
For identical rates, $\mathcal{F}^2_{\mathrm{GHZ}_N}=\tfrac14\big[1+(1-p)^N+p^N\big]+\tfrac12r^N$ and $\mathcal{C}_{\mathrm{GHZ}_N}=e^{-N(\gamma_1/2+\gamma_\varphi)t}$, which for $N=2,3$ gives $\mathcal{F}^2_{\mathrm{GHZ}_2}=\tfrac12\big[1-p+p^2+r^2\big]$ and $\mathcal{F}^2_{\mathrm{GHZ}_3}=\tfrac12\big[1-\tfrac32p+\tfrac32p^2+r^3\big]$.
Under local noise the GHZ coherence decays with an exponent linear in $N$. This is distinct from the quadratic ``superdecoherence'' of Sec.~\ref{sec:multimode_states}, which arises only for collective dephasing.

Comparing Eqs.~\eqref{eq:F_WN} and \eqref{eq:F_GHZN} shows that, for arbitrary local rates, $\mathcal{C}_{\mathrm{GHZ}_2}=\mathcal{C}_{W_2}$ and $\mathcal{C}_{\mathrm{GHZ}_3}=\sqrt{c_{12}c_{13}c_{23}}$, since both sides equal $r_1r_2$ and $r_1r_2r_3$, respectively. Local loss and dephasing therefore give $\mathcal{R}_2=\mathcal{R}_3=1$ in Eqs.~\eqref{eq:chi2} and \eqref{eq:chi3} of the main text.

\subsection{Cluster and general graph states}
\label{sec:cl2}

For a graph state $\ket{G_N}$ on the graph $G=(V,E)$, a closed form valid for arbitrary mode-dependent rates follows from the stabilizer expansion
\begin{equation}
\ket{G_N}\bra{G_N}=\frac{1}{2^N}\sum_{A\subseteq V}\hat g_A,\qquad \hat g_A\equiv\prod_{v\in A}\hat K_v,
\end{equation}
with $\hat K_v$ the stabilizer generators of Eq.~\eqref{eq:cluster_stabilizers}. Up to a sign, $\hat g_A$ acts as $\hat\sigma_x$ or $\hat\sigma_y$ on the modes in $A$, as $\hat\sigma_z$ on the modes in $\mathcal{N}_{\mathrm{odd}}(A)\setminus A$, where $\mathcal{N}_{\mathrm{odd}}(A)$ is the set of vertices with an odd number of neighbors in $A$, and as the identity elsewhere. The single-mode channel $\hat{\Phi}_k$ generated by Eq.~\eqref{eq:master} acts on the Pauli operators as $\hat{\Phi}_k(\hat\sigma_{x,y})=r_k\hat\sigma_{x,y}$, $\hat{\Phi}_k(\hat\sigma_z)=p_k\hat\sigma_z$, and $\hat{\Phi}_k(\mathbb{1})=\mathbb{1}+(1-p_k)\hat\sigma_z$.
The last term generates additional operators $\hat g_A\hat\sigma_z^{(T)}$, with $\hat\sigma_z^{(T)}\equiv\prod_{k\in T}\hat\sigma_z^{(k)}$ and $T$ a nonempty set of modes on which $\hat g_A$ acts trivially. These have the same $\hat\sigma_{x,y}$ support as $\hat g_A$. Since a stabilizer element of a graph state is uniquely determined by this support, they are not in the stabilizer group and have zero expectation value in $\ket{G_N}$. Using $\bra{G_N}\hat g_A\ket{G_N}=1$, we obtain
\begin{equation}
\mathcal{F}^2_{G_N}(t)=\frac{1}{2^N}\sum_{A\subseteq V}\;\prod_{k\in A}r_k(t)\prod_{k\in \mathcal{N}_{\mathrm{odd}}(A)\setminus A}p_k(t).
\label{eq:F_CLN_general}
\end{equation}
Only $A=\varnothing$ survives for $t\to\infty$, giving $\mathcal{F}^2_{G_N}\to2^{-N}=\abs{\bracket{0\dots0}{G_N}}^2$. For pure dephasing ($p_k=1$), Eq.~\eqref{eq:F_CLN_general} reduces to $\mathcal{F}^2_{G_N}=\prod_k(1+r_k)/2$ for any graph.

For the linear chain, $\mathcal{N}_{\mathrm{odd}}(A)\setminus A$ contains the modes outside $A$ with exactly one nearest neighbor in $A$. Summing over the subsets $A$ gives
\begin{align}
\mathcal{F}^2_{\mathrm{CL}^{\mathrm{lin}}_2}(t)&=\tfrac14\Big[1+r_1p_2+p_1r_2+r_1r_2\Big],
\label{eq:F_CL2_asym}\\
\mathcal{F}^2_{\mathrm{CL}^{\mathrm{lin}}_3}(t)&=\tfrac18\Big[1+r_1p_2+p_2r_3+p_1r_2p_3+r_1r_3+r_1r_2p_3+p_1r_2r_3+r_1r_2r_3\Big],
\label{eq:F_CL3_asym}
\end{align}
where, for $N=3$, the terms correspond to $A=\varnothing,\{1\},\{3\},\{2\},\{1,3\},\{1,2\},\{2,3\},\{1,2,3\}$. The expressions are symmetric under the reflection $1\leftrightarrow N$ of the chain. For identical rates, they reduce to $\mathcal{F}^2_{\mathrm{CL}^{\mathrm{lin}}_2}=\tfrac14\big[1+2pr+r^2\big]$ and $\mathcal{F}^2_{\mathrm{CL}^{\mathrm{lin}}_3}=\tfrac18\big[1+2pr+p^2r+r^2+2pr^2+r^3\big]$.
These results illustrate why Eq.~\eqref{eq:dephasing_rule} alone is not sufficient for cluster states: the terms containing $p_k$ originate from population transfer $\ket{1}_k\to\ket{0}_k$ that feeds matrix elements with nonzero overlap with the target.

\subsection{Comparison of the three families}

At short times, Eqs.~\eqref{eq:F_WN}, \eqref{eq:F_GHZN} and \eqref{eq:F_CLN_general} give, with the mode-averaged rates $\bar\gamma_1=\frac1N\sum_k\gamma_1^{(k)}$ and $\bar\gamma_\varphi=\frac1N\sum_k\gamma_\varphi^{(k)}$,
\begin{equation}
\mathcal{F}^2_{W_N}\simeq1-\Big[\bar\gamma_1+\tfrac{2(N-1)}{N}\,\bar\gamma_\varphi\Big]t,
\qquad
\mathcal{F}^2_{\mathrm{GHZ}_N}\simeq \mathcal{F}^2_{G_N}\simeq1-\tfrac{N}{2}\big(\bar\gamma_1+\bar\gamma_\varphi\big)t,
\end{equation}
where the second expression holds for any connected graph state. For $N=2$ all three families decay identically to first order, while for $N\geq3$ the $W$ state has the smallest initial fidelity loss, and the GHZ and connected graph states have the same initial slope.

At long times the ordering changes, because the relaxed vacuum state has overlap $1/2$ with $\ket{\mathrm{GHZ}_N}$ and $2^{-N}$ with $\ket{G_N}$, but none with $\ket{W_N}$: $\mathcal{F}^2_{W_N}\to0$, $\mathcal{F}^2_{\mathrm{GHZ}_N}\to1/2$, and $\mathcal{F}^2_{G_N}\to2^{-N}$. For identical rates and pure amplitude damping, for example, $\mathcal{F}^2_{W_N}$ drops below $\mathcal{F}^2_{\mathrm{GHZ}_N}$ for $\gamma_1t\gtrsim0.74$ ($N=3$) and $\gamma_1t\gtrsim1.02$ ($N=4$). The fidelity at long times is therefore dominated by the vacuum overlap and is not a meaningful measure of the robustness of multipartite coherence, which is instead captured by the coherences $\mathcal{C}(t)$ used in the main text.

Finally, the linear cluster state does not exceed the GHZ state in fidelity, $\mathcal{F}^2_{\mathrm{CL}^{\mathrm{lin}}_N}\leq \mathcal{F}^2_{\mathrm{GHZ}_N}$, in all cases we evaluated ($N=3,4,5$, identical rates with $0\leq\gamma_\varphi/\gamma_1\leq5$, and pure dephasing). For pure dephasing this follows from Jensen's inequality, $\big[(1+r)/2\big]^N\leq(1+r^N)/2$. This fidelity ordering should not be confused with the robustness of the entanglement of cluster states under local measurements (Sec.~\ref{sec:multimode_states}), which is a distinct property.

\clearpage
\newpage
\section{Correlated phase noise and the ratio method}
\label{sec:correlated_noise}

In this section we derive the results used in Sec.~V of the main text: the factorization of phase noise from local decoherence, the coherence decay of $W_N$ and $\mathrm{GHZ}_N$ states under correlated Gaussian phase noise, the ratios $\mathcal{R}_N$, the limiting regimes of purely collective noise, and the analysis of the measured decay times.
Throughout, $\langle\cdot\rangle$ denotes the average over noise realizations $\bm\xi(t)$, $\rho_{\mathbf{mn}}=\bra{\mathbf m}\rho\ket{\mathbf n}$ with $\ket{\mathbf m}=\ket{m_1\ldots m_N}$, $\mathbf d=\mathbf m-\mathbf n$, $\hat N_{\mathrm{tot}}=\sum_k\hat n_k$, and $\lambda_k=\gamma_1^{(k)}/2+\gamma_\varphi^{(k)}$ is the local coherence decay rate of mode $k$ defined in Sec.~\ref{sec:local_decoherence} of these supplementary.

\subsection{Factorization of phase noise and local decoherence}
\label{sec:factorization}

We start from the master equation Eq.~(\ref{eq:mastereq}) of the main text,
$\dot\rho_{\bm\xi}=\mathcal{H}_t[\rho_{\bm\xi}]+\mathcal{L}_{\mathrm{loc}}[\rho_{\bm\xi}]$, with $\mathcal{H}_t[\rho]=-i\big[\sum_k\xi_k(t)\hat n_k,\rho\big]$ and $\mathcal{L}_{\mathrm{loc}}$ the local generator of Eq.~\eqref{eq:master}. The noise $\bm\xi(t)=(\xi_1,\ldots,\xi_N)$ is classical and arbitrary in strength, temporal correlations, and inter-mode correlations.

For a phase vector $\bm\theta$, let $\mathcal{U}_{\bm\theta}[\rho]=e^{-i\bm\theta\cdot\hat{\mathbf n}}\rho\,e^{i\bm\theta\cdot\hat{\mathbf n}}$. Since $e^{i\bm\theta\cdot\hat{\mathbf n}}\hat a_k e^{-i\bm\theta\cdot\hat{\mathbf n}}=e^{-i\theta_k}\hat a_k$, the phase cancels in $\hat a_k\rho\hat a_k^\dagger$ and in $\hat a_k^\dagger\hat a_k$, while $\hat n_k$ commutes with $\bm\theta\cdot\hat{\mathbf n}$. Hence $\mathcal{U}_{\bm\theta}\mathcal{L}_{\mathrm{loc}}\mathcal{U}_{\bm\theta}^{-1}=\mathcal{L}_{\mathrm{loc}}$ for all $\bm\theta$, so that $\mathcal{H}_t$ commutes with $\mathcal{L}_{\mathrm{loc}}$ and with $\mathcal{H}_{t'}$ at all times. For each noise realization,
\begin{equation}
\rho_{\bm\xi}(t)=e^{\mathcal{L}_{\mathrm{loc}}t}\,\mathcal{U}_{\bm\Phi(t)}[\rho(0)],
\qquad
\Phi_k(t)=\int_0^t\xi_k(t')\,\mathrm{d}t'.
\end{equation}
The rotation multiplies $\rho_{\mathbf{mn}}$ by $e^{-i\mathbf d\cdot\bm\Phi}$. The local dynamics only couples matrix elements with the same $\mathbf d$, because $\hat a_k$ lowers $m_k$ and $n_k$ together and $\mathcal{D}[\hat n_k]$ is diagonal in the Fock basis. All elements contributing to $\rho_{\mathbf{mn}}(t)$ therefore carry the same phase factor, and averaging over the noise gives
\begin{equation}
\rho_{\mathbf{mn}}(t)=\big\langle e^{-i\mathbf d\cdot\bm\Phi(t)}\big\rangle\,\rho^{(0)}_{\mathbf{mn}}(t),
\qquad
\rho^{(0)}(t)=e^{\mathcal{L}_{\mathrm{loc}}t}[\rho(0)],
\label{eq:factorization_lemma}
\end{equation}
where $\rho^{(0)}(t)$ denotes the solution of Eq.~(\ref{eq:mastereq}) with $\bm\xi=0$. This result holds for arbitrary local rates and arbitrary noise statistics.

\subsection{Coherences of $W_N$ and $\mathrm{GHZ}_N$ states}
\label{sec:coh_WGHZ}

\textbf{Local factor.} For every coherence of $\ket{W_N}$ ($\mathbf d=\mathbf e_k-\mathbf e_l$, with $\mathbf e_k$ the $k$th unit vector) and of $\ket{\mathrm{GHZ}_N}$ ($\mathbf d=(1,\ldots,1)$), each mode is excited in at most one of the two Fock states, i.e.\ $m_k+n_k=d_k^2$. Moreover, no element with the same $\mathbf d$ and higher occupation is populated, so the equation of motion is closed (Sec.~\ref{sec:local_decoherence}) and
\begin{equation}
\rho^{(0)}_{\mathbf{mn}}(t)=e^{-\sum_kd_k^2\lambda_kt}\,\rho_{\mathbf{mn}}(0).
\end{equation}

\textbf{Noise factor.} For Gaussian phase noise with covariance matrix $\mathcal{K}_{kl}(t)=\langle\delta\Phi_k(t)\,\delta\Phi_l(t)\rangle$, $\delta\Phi_k=\Phi_k-\langle\Phi_k\rangle$, one has $\big|\langle e^{-i\mathbf d\cdot\bm\Phi}\rangle\big|=\exp\big[-\tfrac12\sum_{k,l}d_kd_l\mathcal{K}_{kl}\big]$. Deterministic phases $\langle\Phi_k\rangle$ only rotate the coherences and drop out of their absolute values. Combining both factors,
\begin{equation}
\frac{\abs{\rho_{\mathbf{mn}}(t)}}{\abs{\rho_{\mathbf{mn}}(0)}}
=\exp\Big[-\sum_kd_k^2\lambda_kt-\frac12\sum_{k,l}d_kd_l\,\mathcal{K}_{kl}(t)\Big].
\label{eq:coh_general}
\end{equation}
For the pairwise $W_N$ coherences and the $\mathrm{GHZ}_N$ coherence this gives
\begin{align}
c_{kl}(t)&=\exp\Big[-(\lambda_k+\lambda_l)\,t-\tfrac12\big(\mathcal{K}_{kk}+\mathcal{K}_{ll}-2\mathcal{K}_{kl}\big)\Big],
\label{eq:ckl_general}\\
\mathcal{C}_{\mathrm{GHZ}_N}(t)&=\exp\Big[-\sum_k\lambda_k\,t-\tfrac12\sum_k\mathcal{K}_{kk}-\sum_{k<l}\mathcal{K}_{kl}\Big].
\label{eq:CGHZ_general}
\end{align}
The $W_N$ coherences depend only on the phase differences $\Phi_k-\Phi_l$, the GHZ coherence on the total phase $\sum_k\Phi_k$.
The total coherence of Eq.~(\ref{eq:coherence}) of the main text is $\mathcal{C}_{W_2}=c_{12}$ for $N=2$. For $N=3$ it is the mean of the $c_{kl}$ weighted by $\abs{\rho_{1_k,1_l}(0)}$, which reduces to $\frac13\sum_{k<l}c_{kl}$ for equal initial coherences.

\subsection{Ratios $\mathcal{R}_N$}
\label{sec:ratio_method}

In the product of all $\binom{N}{2}$ pairwise coherences, each local rate $\lambda_k$ and each single-mode variance $\mathcal{K}_{kk}$ appears exactly $N-1$ times. Therefore
\begin{equation}
\mathcal{R}_N(t)\equiv\frac{\mathcal{C}_{\mathrm{GHZ}_N}(t)}{\big[\prod_{k<l}c_{kl}(t)\big]^{1/(N-1)}}
=\exp\Big[-\frac{N}{N-1}\sum_{k<l}\mathcal{K}_{kl}(t)\Big]
\label{eq:RN_general}
\end{equation}
is independent of all local loss and dephasing rates and of all single-mode phase variances. For $N=2,3$ it reduces to Eqs.~(\ref{eq:chi2}) and (\ref{eq:chi3}) of the main text. Two limits are relevant:
\begin{enumerate}
\item \emph{Independent noise.} If the $\xi_k$ are mutually independent, $\langle e^{-i\mathbf d\cdot\bm\Phi}\rangle=\prod_k\langle e^{-id_k\Phi_k}\rangle$. Since $\abs{\langle e^{-i\Phi_k}\rangle}=\abs{\langle e^{+i\Phi_k}\rangle}$, each mode contributes the same factor to numerator and denominator, and $\mathcal{R}_N=1$ for arbitrary, not necessarily Gaussian, noise statistics and spectra.
\item \emph{Collective noise.} If $\xi_k(t)=\xi(t)$ for all $k$, all phases equal the common phase $\Phi(t)=\int_0^t\xi(t')\,\mathrm{d}t'$, all entries of $\mathcal{K}$ equal $\langle\delta\Phi(t)^2\rangle$, and $\mathcal{R}_N(t)=\exp[-N^2\langle\delta\Phi(t)^2\rangle/2]$, the GHZ superdecoherence factor discussed in Sec.~\ref{sec:collective}.
\end{enumerate}
If the covariances grow linearly in time, $\mathcal{K}_{kl}(t)=2D_{kl}t$, the symmetric matrix $D$ acts as a dephasing-rate matrix whose off-diagonal elements are the correlated dephasing rates. Equation~\eqref{eq:RN_general} then gives $\mathcal{R}_N=e^{-\Gamma_Nt}$ with
\begin{equation}
\Gamma_N=\frac{2N}{N-1}\sum_{k<l}D_{kl},
\qquad\text{i.e.}\qquad
\Gamma_2=4D_{12},\quad \Gamma_3=3\sum_{k<l}D_{kl}.
\label{eq:GammaN_D}
\end{equation}
Note that $D_{k\neq l}$ can be negative for anticorrelated noise.
With $\hat Z_k=1-2\hat n_k$, $D$ coincides with $\gamma C$, where $\gamma$ and $C$ are the overall dephasing rate and the dephasing coefficient matrix of Ref.~\cite{Wang2026}. For stationary noise, $D_{kl}=S_{kl}(0)/2$, with $S_{kl}(\omega)=\int_{-\infty}^{\infty}\mathrm{d}\tau\,e^{i\omega\tau}\langle\delta\xi_k(\tau)\,\delta\xi_l(0)\rangle$ the cross-spectral density of the mode-frequency fluctuations~\cite{PazSilva2017}.

\subsection{Purely collective noise}
\label{sec:collective}

For collective noise, $\hat H_{\mathrm{noise}}=\xi(t)\hat N_{\mathrm{tot}}$ (with $\hbar=1$), we take $\xi(t)$ to be a zero-mean, stationary Gaussian process with autocorrelation function $R(\tau)=\langle\xi(t)\xi(t+\tau)\rangle$, variance $\sigma^2=R(0)$, and accumulated phase $\Phi(t)=\int_0^t\xi(t')\,\mathrm{d}t'$. A coherence whose total excitation numbers differ by $\Delta n=\sum_kd_k$ then decays as
\begin{equation}
\frac{\rho_{\mathbf{mn}}(t)}{\rho^{(0)}_{\mathbf{mn}}(t)}=\exp\Big[-\frac{\Delta n^2}{2}\langle\Phi(t)^2\rangle\Big],
\qquad
\langle\Phi(t)^2\rangle=\int_0^t\!\!\mathrm{d}t_1\!\int_0^t\!\!\mathrm{d}t_2\,R(t_1-t_2)=2\int_0^t\mathrm{d}\tau\,(t-\tau)R(\tau).
\label{eq:rhonm_collective}
\end{equation}
For exponentially correlated (Ornstein--Uhlenbeck) noise, $R(\tau)=\sigma^2e^{-\nu\abs{\tau}}$ with correlation time $1/\nu$,
\begin{equation}
\langle\Phi(t)^2\rangle=\frac{2\sigma^2}{\nu^2}\big(\nu t-1+e^{-\nu t}\big).
\label{eq:PhiVarianceOU}
\end{equation}
Differentiating Eq.~\eqref{eq:rhonm_collective} and using $[\hat N_{\mathrm{tot}},[\hat N_{\mathrm{tot}},\rho]]_{\mathbf{mn}}=\Delta n^2\rho_{\mathbf{mn}}$ gives the time-convolutionless master equation
\begin{equation}
\dot\rho=\mathcal{L}_{\mathrm{loc}}\rho-\Big(\int_0^tR(\tau)\,\mathrm{d}\tau\Big)\big[\hat N_{\mathrm{tot}},[\hat N_{\mathrm{tot}},\rho]\big].
\label{eq:NonMarkovME}
\end{equation}
It is exact because the noise Hamiltonians at different times commute and the cumulant expansion of Gaussian noise terminates at second order.

\vspace{1mm}
\textbf{Markovian regime} ($\nu t\gg1$). The rate in Eq.~\eqref{eq:NonMarkovME} saturates, giving the Lindblad equation
\begin{equation}
%\dot\rho=\Gamma_c\,\mathcal{D}[\hat N_{\mathrm{tot}}]\rho,
\dot\rho=\mathcal{L}_{\mathrm{loc}}\rho+\Gamma_c\,\mathcal{D}[\hat N_{\mathrm{tot}}]\rho,
\qquad
\Gamma_c=2\int_0^\infty R(\tau)\,\mathrm{d}\tau=\frac{2\sigma^2}{\nu},
\label{eq:MarkovME}
\end{equation}
and $\langle\Phi(t)^2\rangle\simeq\Gamma_c(t-1/\nu)$. Coherences decay at rate $\Delta n^2\Gamma_c/2$. This defines the collective-dephasing rate $\Gamma_c$ of the main text, $\langle\delta\Phi(t)^2\rangle=\Gamma_ct$, up to the constant offset $-\Gamma_c/\nu$. A single-mode coherence ($\Delta n=1$) decays at $\Gamma_c/2$, which is the contribution of collective noise to the pure-dephasing rate of each mode.

\vspace{1mm}
\textbf{Quasi-static regime} ($\nu t\ll1$). The detuning is constant within a shot and varies only between shots, so that $\langle\Phi(t)^2\rangle\simeq\sigma^2t^2$. Coherences decay as $\exp[-\Delta n^2\sigma^2t^2/2]$, a Gaussian decay that no time-independent Lindblad equation can describe.

\vspace{1mm}
All coherences of $\ket{W_N}$ have $\Delta n=0$, so $W$ states are decoherence-free under collective noise of any spectrum. More generally, the same holds for any state supported on a single eigenspace of $\hat N_{\mathrm{tot}}$. The GHZ coherence has $\abs{\Delta n}=N$, the largest value in the computational subspace, and acquires the factor $e^{-N^2\langle\Phi(t)^2\rangle/2}$ on top of the local decay. In the Markovian regime this is $e^{-N^2\Gamma_ct/2}$, while in the quasi-static regime it is a Gaussian with $1/e$ time $\sqrt2/(N\sigma)$.
For comparison, independent dephasing with rate $\gamma_\varphi$ per mode gives $e^{-N\gamma_\varphi t}$, whereas collective noise with the same single-mode rate, $\Gamma_c/2=\gamma_\varphi$, gives $e^{-N^2\gamma_\varphi t}$, the GHZ superdecoherence.
Since the pairwise $W_N$ coherences are unaffected, Eq.~\eqref{eq:RN_general} reproduces $\mathcal{R}_N=e^{-N^2\langle\Phi^2\rangle/2}$, i.e. $\Gamma_N=N^2\Gamma_c/2$ and, from Eq.~\eqref{eq:GammaN_D}, $D_{kl}=\Gamma_c/2$ for all $k,l$.

\subsection{Analysis of the measured decay times}
\label{sec:ratio_analysis}

Each normalized coherence ($\mathcal{C}_{W_2}$, $\mathcal{C}_{W_3}$, $c_{kl}$, $\mathcal{C}_{\mathrm{GHZ}_2}$, $\mathcal{C}_{\mathrm{GHZ}_3}$) is fitted with $A\,e^{-t/\tau}+B$. The offset $B$ accounts for the positive bias of $\abs{\rho_{\mathbf{mn}}}$ estimated from noisy tomographic data, which leaves a residual floor at long delays. Only the decay times $\tau$, reported with their one-standard-deviation fit uncertainties $\sigma_\tau$ in Fig.~\ref{fig3} of the main text, enter the analysis below. Normalizing to $t=0$ removes state-preparation and measurement errors, which are identical for all delays.

Within the exponential description, Eq.~\eqref{eq:RN_general} relates the decay rates of numerator and denominator,
\begin{equation}
\Gamma_2=\frac{1}{\tau_{\mathrm{GHZ}_2}}-\frac{1}{\tau_{W_2}},
\qquad
\Gamma_3=\frac{1}{\tau_{\mathrm{GHZ}_3}}-\frac12\sum_{k<l}\frac{1}{\tau_{kl}},
\label{eq:GammaN_tau}
\end{equation}
with $\tau_{kl}$ the decay time of $c_{kl}$. Assuming independent fit errors, we propagate uncertainties as
\begin{equation}
\sigma_{\Gamma_2}^2=\frac{\sigma_{\tau_{\mathrm{GHZ}_2}}^2}{\tau_{\mathrm{GHZ}_2}^4}+\frac{\sigma_{\tau_{W_2}}^2}{\tau_{W_2}^4},
\qquad
\sigma_{\Gamma_3}^2=\frac{\sigma_{\tau_{\mathrm{GHZ}_3}}^2}{\tau_{\mathrm{GHZ}_3}^4}+\frac14\sum_{k<l}\frac{\sigma_{\tau_{kl}}^2}{\tau_{kl}^4}.
\end{equation}
The three $\tau_{kl}$ are extracted from the same set of density matrices, so treating their errors as independent is an approximation.

The second term of $\Gamma_3$ is the purely local GHZ$_3$ decay rate predicted from the pairwise coherences, without free parameters. The corresponding lifetime is $\tau_{\mathrm{loc}}=\big(\frac12\sum_{k<l}\tau_{kl}^{-1}\big)^{-1}$, and the fraction of the measured GHZ$_3$ decay rate it accounts for is $\tau_{\mathrm{GHZ}_3}/\tau_{\mathrm{loc}}$.

The initial slope of $\mathcal{C}_{W_3}$ is the (weighted) mean of the three pairwise rates. Agreement between $1/\tau_{W_3}$ and $\frac13\sum_{k<l}1/\tau_{kl}$ therefore checks the pairwise fits. Note that $\tau_{W_3}$ itself is not used in Eq.~\eqref{eq:GammaN_tau}.

Upper limits on a quantity $X$ with estimate $\bar X$ and uncertainty $\sigma_X$ are one-sided at $95\,\%$ confidence, $X<\bar X+1.645\,\sigma_X$. Using Eq.~\eqref{eq:GammaN_D}, we bound $D_{12}=\Gamma_2/4$ and $\sum_{k<l}D_{kl}=\Gamma_3/3$; these require no assumption on the spatial structure of the noise. 
For purely collective noise, each data set gives an estimate $\Gamma_c^{(N)}=2\Gamma_N/N^2$, i.e.\ $\Gamma_c^{(2)}=\Gamma_2/2$ and $\Gamma_c^{(3)}=2\Gamma_3/9$, with $\sigma_{\Gamma_c}^{(N)}=2\sigma_{\Gamma_N}/N^2$.
The $N=2$ and $N=3$ estimates come from independent data sets, so we combine them into the inverse-variance weighted mean
\begin{equation}
\bar\Gamma_c=\frac{\sum_Nw_N\Gamma_c^{(N)}}{\sum_Nw_N},
\qquad
w_N=\big(\sigma_{\Gamma_c}^{(N)}\big)^{-2},
\qquad
\sigma_{\bar\Gamma_c}=\Big(\sum_Nw_N\Big)^{-1/2}.
\end{equation}
Individual estimates are not truncated at the physical boundary $\Gamma_c\geq0$ before averaging, to avoid biasing the mean. The resulting values are summarized in Table~\ref{tab:corrnoise} of the main text.

A bound on $\Gamma_c$ translates into a bound $\Gamma_c/2$ on the collective contribution to the pure-dephasing rate of each mode. Unlike an estimate from single-mode Ramsey measurements, which requires subtracting $1/(2T_{1,k})$ from $1/T^\ast_{2,k}$ in order to obtain the local pure-dephasing rate, this bound does not depend on the calibration of the single-mode relaxation times, since all local contributions cancel in $\mathcal{R}_N$.

\end{document}